\documentclass[aps,twocolumn,notitlepage,pra,superscriptaddress,nofootinbib]{revtex4-2}

\usepackage[dvipsnames]{xcolor}
\usepackage{hyperref}
\hypersetup{
  breaklinks=true,
  colorlinks=true,
  allcolors=BlueViolet,
}

\usepackage{amsmath,amssymb,bm}
\usepackage{graphicx}
\usepackage{epstopdf}
\usepackage{latexsym}
\usepackage{enumitem}
\usepackage{setspace}
\usepackage[normalem]{ulem}
\usepackage{natbib}
\usepackage{braket}
\usepackage{comment}
\usepackage{soul}
\usepackage{listings}

\definecolor{dkgreen}{rgb}{0,0.6,0}
\definecolor{gray}{rgb}{0.5,0.5,0.5}
\definecolor{mauve}{rgb}{0.58,0,0.82}

\newcommand{\Rev}[1]{{\color{black}{#1}\normalcolor}} % Revision
\begin{document}

\title{Variational quantum state preparation within an entangle-rotate circuit framework for quantum-enhanced metrology in noisy systems}
\author{Juan C. Zu\~{n}iga Castro}
\affiliation{Homer L. Dodge Department of Physics and Astronomy, The University of Oklahoma, Norman, OK 73019, USA}
\affiliation{Center for Quantum Research and Technology, The University of Oklahoma, Norman, OK 73019, USA}
\author{Jeffrey Larson}
\author{Matt Menickelly}
\author{Sri Hari Krishna Narayanan}
\affiliation{Mathematics and Computer Science Division, Argonne National Laboratory,  Lemont, IL 60439, USA}
\author{Yicheng Zhang}
\affiliation{Homer L. Dodge Department of Physics and Astronomy, The University of Oklahoma, Norman, OK 73019, USA}
\affiliation{Center for Quantum Research and Technology, The University of Oklahoma, Norman, OK 73019, USA}
\affiliation{Department of Physics and Astronomy, George Mason University, Fairfax, Virginia 22030, USA}
\author{Michael A. Perlin}
\affiliation{Global Technology Applied Research, JPMorgan Chase, New York, NY 10017, USA}
\affiliation{Infleqtion, Inc., 
%141 West Jackson Blvd Suite 1875 
Chicago, IL 60604, USA}
\author{Robert J. Lewis-Swan}
\affiliation{Homer L. Dodge Department of Physics and Astronomy, The University of Oklahoma, Norman, OK 73019, USA}
\affiliation{Center for Quantum Research and Technology, The University of Oklahoma, Norman, OK 73019, USA}
\date{\today}

\begin{abstract}
  We investigate the generation of quantum states for precision metrology in noisy two-level systems. These states are obtained by optimizing a variational quantum circuit to maximize the quantum Fisher information (QFI) of the output state for a given decoherence rate and interaction Hamiltonian. The circuit architecture, inspired by twist-and-turn schemes, features a sequence of $n$ entangling layers, each consisting of entangling gates followed by a global rotation. We observe notable improvements in the QFI as the circuit layer depth increases, even for appreciable noise rates, demonstrating that our entangle-rotate architecture expands the accessible state space under realistic noise conditions. Our approach thus provides a general and efficient framework for generating quantum-enhanced sensing states. Our analysis extends to systems of power-law interactions spanning from all-to-all to nearest-neighbor interactions. We also analyze the capabilities of our circuit to prepare states for system sizes greater than $8$ qubits.
  \end{abstract}

\maketitle
\section{Introduction}\label{sec:introduction}
The use of entangled quantum states for enhanced parameter estimation is a topic of intensive research across various areas of modern physics~\cite{ligoo4detectorcollaboration2023broadband, RevModPhys.90.035005, quantum-spin-squeezing, optical_clock_transitions, GilmoreKevinA.2021Qsod}. In a typical estimation protocol, the precision with which a parameter $\phi$ can be estimated with a single measurement is fundamentally limited by the quantum Cram\'{e}r--Rao bound $(\Delta\phi)^2\geq1/F_Q$, where $F_Q$ denotes the quantum Fisher information of the probe state. However, experimentally realistic sensing scenarios are typically affected by decoherence and technical noise, which degrade the quality of probe states, specifically their quantum Fisher information (QFI), and consequently reduce readout sensitivities. Thus, as contemporary quantum hardware remains far from ideal~\cite{Jiao_2023}, optimal state preparation in noisy environments remains an open challenge, making it crucial to approach sensing protocols pragmatically by formulating them within an open-system framework. 

\Rev{Variational approaches have emerged as promising tools for the identification and use of metrologically useful quantum states in both closed- and open-system settings~\cite{npj-dipolar, photonic-variational, PhysRevLett.131.073602, VQE-Bravo-PrietoCarlos2020Sovq, zoller-vqc-ramsey-intfrmtry, MacLellan2024variational, qfi_opt_manuscript1, Koczor_2020, PhysRevLett.128.160505, PRXQuantum.4.020333, Liu2025Variational}, especially as the latter approach typically becomes intractable with increasing system size~\cite{Koczor_2020}.} In this vein, in Ref.~\cite{qfi_opt_manuscript1} we investigated the use of a low-depth variational quantum circuit (VQC) to prepare optimal probe states under noisy conditions. While we identified strategies for generating states with appreciable QFIs, we found that the VQC's restrictive architecture limited access to the full Hilbert space~\cite{LaroccaMartín2023Tooi} and consequently the preparation of states with larger QFI. Notably, we observed that similar Hamiltonians yielded significantly different performances in generating useful states, even under identical initial conditions, and that increasing decoherence rates consistently degraded the VQC's ability to generate useful entanglement. These findings highlight the need to modify our ansatz to improve its capacity to explore a larger portion of the Hilbert space and to enhance the ansatz's robustness against noise.

A promising method to increase state resilience to noise is to accelerate the generation of entanglement via so-called twist-and-turn schemes in which (i) entanglement is typically generated by a Hamiltonian that is straightforward to implement and (ii) the system is stroboscopically propagated by an auxiliary drive~\cite{TnT-Oberthaler, zoller-vqc-ramsey-intfrmtry, LiuYC2011Ssto, holland_sq, CarrascoSebastianC.2022ESSv, TnT-Thurtell, TnT-IBR-Haine, exponential_entanglment_TAT}. Such strategies have been shown to emulate Hamiltonians that feature rapid entanglement generation and have the capacity to yield states of metrological utility. Because decoherence has less time to act during state formation, faster entanglement generation can lead to the more robust preparation of complex entangled states~\cite{holland_sq, yicheng_dicke_2025}. Moreover, Hamiltonian emulation via the use of an auxiliary drive can expand the accessible dynamics, thereby enabling the preparation of states inaccessible to the native Hamiltonian~\cite{LloydSeth1996UQS, gietka2021simulating}. 

In this work we modify the ansatz of Ref.~\cite{qfi_opt_manuscript1}, enhancing its flexibility by replacing its single entangling gate with a layered entangle--rotate (E--R) structure, which emulates the alternating dynamics of twist-and-turn protocols. This design enables the preparation of a broader class of states with appreciable QFI for a given Hamiltonian. Our results demonstrate clear QFI enhancements as well as reductions in cumulative interaction duration \Rev{for increasing circuit depth in identified parameter regimes.} \Rev{Beyond the increased parameter space given by additional layers, a major advance of this work over Ref.~\cite{qfi_opt_manuscript1} is that the intermediate rotations allow us to break a redundancy coming from repeated application of a fixed entangling generator. Because the entangling Hamiltonians and the dissipative Lindbladian generator are both time-independent in the entangling epochs, consecutive entangling epochs with no intervening operation compose additively in time and are no more expressive than a single epoch of adjusted duration. By reorienting the state, we can access parts of Hilbert space that are unreachable under the native, single-generator dynamics of Ref.~\cite{qfi_opt_manuscript1}. This mechanism is conceptually analogous to the use of intermediate quantum kicks to connect dynamics generated by distinct Hamiltonians in Ref.~\cite{gietka2021simulating}.}

We present our findings in the following order. In Sec.~\ref{sec:vqc}, we introduce our VQC, outlining the sequence of fundamental entangling and global operations. In Sec.~\ref{sec:entangling-dynamics}, we present the entangling mechanisms and sources of decoherence applied during the VQC's entangling stages. We also describe our optimization methodologies, which employ the QFI as the figure of merit. In Sec.~\ref{sec:results}, we present optimal state preparations for sensing global spin rotations for various decoherence strengths. In Sec.~\ref{sec:interaction-range} we analyze our VQC's state generation ability for different interaction ranges. In Sec.~\ref{sec:system-size}, we present results for state generation in larger system sizes. Section~\ref{sec:conclusion} concludes and briefly discusses future research directions.

\section{Variational quantum state preparation}\label{sec:vqc}
\begin{figure}
    \centering
     \includegraphics[width=8.6cm]{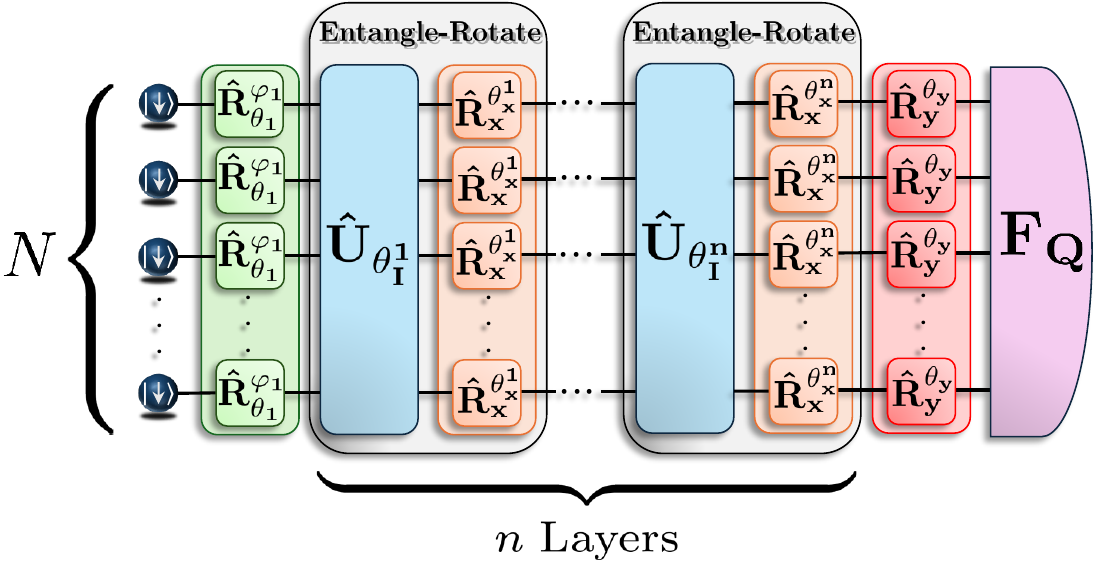}
    \caption{A one-dimensional chain of $N$ qubits initialized in the state $\ket{\downarrow}^{\otimes N}$ is subjected to a VQC (left to right). The qubits are globally rotated by $\hat{R}_{\theta_1}^{\varphi_1}$ through an angle $\theta_1$ about an axis defined by the azimuthal angle $\varphi_1$. The state is then subjected to a set of $n$ entangling gates $\hat{U}_{\theta^{k}_{\mathrm{I}}}$ defined by characteristic strength $\theta_{\mathrm{I}}^{k} = \chi t_{\mathrm{I}}^{(k)}$, where $k\in\{1, \cdots, n\}$ and $n$ is the number of entangle-rotate layers. Each entangling gate is followed by a global $x$ rotation $\hat{R}_x^{\theta_x^{k}}$ by the angle $\theta_x^{k}$. Finally, the qubits are globally rotated by a $y$ rotation $\hat{R}_y^{\theta_y}$ for an angle $\theta_y$. We compute the QFI ($F_Q$) of the prepared state, $\hat{\rho}(\mathbf{x}^{(n)})$, which is parameterized by $\mathbf{x}^{(n)} = (\theta_1, \varphi_1, \theta^{1}_{\mathrm{I}}, \theta_x^{1}, \cdots, \theta^{n}_{\mathrm{I}}, \theta_x^{n}, \theta_y)$. See the main text for a detailed definition of the VQC stages and associated parameters.}
    \label{fig:VQC}
\end{figure}

Our investigation uses a variable-depth variational quantum circuit to prepare quantum states of $N$ interacting qubits to detect global qubit rotations. The VQC architecture is composed of a three-stage process that prepares the quantum state for a sensing task. The state is first rotated in preparation for a sequence of $n$ E--R layers. Within each E--R layer, the state undergoes a period of entanglement generation via two-qubit interactions that is followed by an $x$-axis rotation. After the sequence of E--R layers, the protocol concludes with a global $y$-axis rotation. 

The VQC sequence (see Fig.~\ref{fig:VQC}) begins by initializing an ensemble of $N$ qubits in a pure product state $\ket{\psi_0} = \ket{ \downarrow }^{\otimes N}$ in the $z$ basis. The initial rotation is parameterized as $\hat{R}_{\theta_1}^{\varphi_1} = \prod_{j=1}^{N} e^{-i\theta_1\hat{\sigma}^{\varphi_1}_j/2}$, where $\hat{\sigma}_j^{\varphi_1} = \hat{\sigma}^x_j\cos{(\varphi_1)} + \hat{\sigma}_j^y\sin{(\varphi_1)}$ is constructed from Pauli matrices $\hat{\sigma}_j^{\beta}$ with $\beta\in x,y,z$. In $\hat{R}_{\theta_1}^{\varphi_1}$, $\theta_1$ and $\varphi_1$ define the rotation angle and axis, respectively. Each entangling process applied thereafter is generated by the gate: 
\begin{equation}\label{eqn:general-entangling-gate}
    \hat{U}_{\theta_{\mathrm{I}}^{k}} = e^{-it^{(k)}_{\mathrm{I}}\hat{H}_{\mathrm{I}}(\chi)},
\end{equation}
where the Hamiltonian $\hat{H}_{\mathrm{I}}(\chi)$ describes the qubit-qubit interactions, $\chi$ denotes the characteristic coupling strength and we set $\hbar = 1$. For a set of $n$ E--R layers, the ``strength'' of the $k$th entangling gate is set by $\theta_{\mathrm{I}}^k = \chi t_{\mathrm{I}}^{(k)}$, where $t_{\mathrm{I}}^{(k)}$ is the gate duration and where $k \in \{1, . . ., n\}$. Although we initially present the entangling gate as a unitary operation, the role of decoherence will shortly be incorporated into the state preparation dynamics. Furthermore, the $x$-axis rotations following each entangling process are given by $\hat{R}_x^{\theta^k_x} = e^{-i\theta_x^k \hat{S}_x}$, where we define the collective operator $\hat{S}_{\beta} = \sum_{j=1}^{N}{\hat{\sigma}_j^{\beta}/2}$. The final rotation is similarly $\hat{R}_y^{\theta_y} = e^{-i\theta_y \hat{S}_y}$.

The VQC is thus characterized by a total of $2n + 3$ parameters, collectively denoted by $\mathbf{x}^{(n)} = (\theta_1, \varphi_1, \theta^{1}_{\mathrm{I}}, \theta_x^{1}, \cdots, \theta^{n}_{\mathrm{I}}, \theta_x^{n}, \theta_y)$, and by the choice of entangling Hamiltonian, $\hat{H}_{\mathrm{I}}(\chi)$. We denote the output state of the VQC in terms of these parameters by means of the density matrix $\hat{\rho}(\mathbf{x}^{(n)})$.

\section{Entangling dynamics}\label{sec:entangling-dynamics}
We implement interactions via the Ising and finite-range two-axis twisting (FTAT) Hamiltonians $\hat{H}_{\mathrm{I}}(\chi)$ characterizing the entangling gate:
\begin{align} \label{eqn:ising-ham} 
\hat{H}_{\mathrm{Ising}}(\chi) &= \sum_{i\ne j}{\chi_{ij} \hat{\sigma}_i^z\hat{\sigma}_j^z} \text{ and}\\ 
    \label{eqn:tat-ham}
    \hat{H}_{\text{FTAT}}(\chi) &= \sum_{i\ne j}{\chi_{ij}(\hat{\sigma}_i^x\hat{\sigma}_j^y + \hat{\sigma}_i^y\hat{\sigma}_j^x)/2}.
\end{align} 
Here $i$ and $j$ index individual qubits at positions $r_i$ and $r_j$, respectively, and the couplings $\chi_{ij} = \chi/\mathcal{N}_{\alpha}\vert r_i-r_j\vert^{\alpha}$, where $\mathcal{N}_{\alpha} = (N-1)^{-1}\sum_{i\neq j} \vert r_i - r_j\vert^{-\alpha}$ is a normalization factor to ensure that the Hamiltonian is extensive~\cite{kac,kacDefenuNicolo2021Mads} and to facilitate comparisons between the models as a function of the decoherence strength. We use $\alpha$ to set the interaction range. Our work focuses on qubits in a one-dimensional geometry, which can be realized by using, for example, arrays of regularly spaced neutral atoms or trapped ions~\cite{Foss-FeigM2013Dqco-ising-cat-states, PhysRevLett.92.207901}.

We streamline our VQC by including decoherence only during the entangling blocks and approximating the global rotations as unitary, consistent with operating regimes where rotation timescales are short compared with the interaction stages. The entangling processes of Eq.~\eqref{eqn:general-entangling-gate} are thus modeled by using the Lindblad master equation, 
\begin{equation}\label{eqn:master-eqn}
\frac{d\hat{\rho}}{dt} = -i[\hat{H}_{\mathrm{I}}(\chi), \hat{\rho}] + \sum_{\nu, j} \gamma_{\nu}\left(\hat{L}^{\nu}_j\hat{\rho}\hat{L}^{\nu\dagger}_j - \frac{1}{2}\{\hat{L}^{\nu\dagger}_j\hat{L}^{\nu}_j, \hat{\rho}\}\right),
\end{equation}
where $[\hat{A},\hat{B}]=\hat{A}\hat{B}-\hat{B}\hat{A}$ and $\{\hat{A}, \hat{B}\} = \hat{A}\hat{B} + \hat{B}\hat{A}$ respectively denote the commutator and anticommutator of operators $\hat{A}$ and $\hat{B}$ and where $\hat{H}_\mathrm{I}(\chi)$ is the chosen entangling Hamiltonian [either Eq.~\eqref{eqn:ising-ham} or Eq.~\eqref{eqn:tat-ham}]. The jump operators $\hat{L}^{\nu}_{j}$ describe decoherence acting on the $j$th qubit at a rate $\gamma_{\nu}$. Throughout, we focus on axis-dependent dephasing with $\hat{L}^{\nu}_{j} = \hat{\sigma}_j^{\nu}/2$ for $\nu \in\set{x, \:y, \:z}$. Specifically, we subject our interaction stages to equivalent decoherence rates across all axes, that is, we set $\gamma_x = \gamma_y = \gamma_z = \gamma$. In this way, we remove any possible axis biases inherent in the Hamiltonians of Eqs.~(\ref{eqn:ising-ham}) and~(\ref{eqn:tat-ham}).

\subsection{Optimization of the quantum Fisher information}
We seek to identify optimal state preparation protocols, within the constraints of the VQC defined above, for detecting global spin rotations. The quantity capturing the suitability of a generic quantum state $\hat{\rho}$ for this task is the quantum Fisher information~\cite{Brun_2014_metrology, qfi-toth},
\begin{equation}\label{eqn:qfi}
    F_Q[\hat{\rho}(\mathbf{x}^{(n)}) ; \hat{S}_z] = 2\sum_{i, j}\frac{(\lambda_i - \lambda_j)^2}{\lambda_i + \lambda_j}\vert\braket{i\vert \hat{S}_z\vert j}\vert^2. 
\end{equation}
Here, $\ket{j}$ and $\lambda_j$ correspond, respectively, to the eigenvectors and eigenvalues of the density matrix $\hat{\rho}(\mathbf{x}^{(n)})$, that is, $\hat{\rho} = \sum_{j} \lambda_j \ket{j}\!\bra{j}$, and $\hat{S}_z$ is the generator of the unitary transformation $\hat{U}_{\phi} = e^{-i\phi\hat{S}_z}$ encoding a classical parameter $\phi$. The QFI, via the quantum Cram\'{e}r--Rao bound, places a bound on the minimum uncertainty with which the classical parameter $\phi$ can be estimated, $(\Delta\phi)^2 \geq 1/F_Q[\hat{\rho}(\mathbf{x}^{(n)}) ; \hat{S}_z]$.

In the following, we refer to an uncorrelated product state as a state that enables estimation uncertainty at the standard quantum limit (SQL), $(\Delta\phi)^2 \ge 1/N$ (equivalently, $F_Q[\hat{\rho}(\mathbf{x}^{(n)}) ; \hat{S}_z] \le N$). A typical example of an uncorrelated product state is a coherent spin state wherein the qubits are collectively polarized along a common axis, that is, $\ket{\rightarrow}^{\otimes N}$ with $\ket\rightarrow \propto \ket\uparrow + \ket\downarrow$. The introduction of correlations and entanglement can lead to quantum-enhanced estimation below the SQL, but estimation remains fundamentally bounded by the Heisenberg limit (HL), $(\Delta\phi)^2 \ge 1/N^2$ (equivalently, $F_Q[\hat{\rho}(\mathbf{x}^{(n)}) ; \hat{S}_z] \le N^2$). It is known that an optimal state for sensing rotations about $\hat{z}$ that saturates the HL is the macroscopic superposition (``cat'' or GHZ) state~\cite{qfi-toth},
\begin{equation}\label{eqn:ghz-state}
    \ket{\mathrm{GHZ},\Phi} = (\ket{\uparrow}^{\otimes N} + e^{i\Phi} \ket{\downarrow}^{\otimes N})/\sqrt{2}, 
\end{equation}
where $\Phi$ denotes an arbitrary phase.

In Sec.~\ref{sec:results} we take into account the effects of decoherence and investigate optimal preparation strategies using our E--R-based approach.
For a fixed decoherence strength $\gamma$, we obtain the state $\hat{\rho}(\mathbf{x}^{(n)})$ generated by a VQC of $n$ E--R layers with parameters $\mathbf{x}^{(n)}$. We numerically integrate the Lindblad master equation (see Eq.~\eqref{eqn:master-eqn}) within each E--R layer's entangling gate $\hat{U}_{\theta_{\mathrm{I}}^{k}}$ with choice of entangling Hamiltonian $\hat{H}_{\mathrm{I}}(\chi)$.

\subsection{Optimization methods}
In the following, we separate our numerical formulation into two cases: finite-range interactions ($\alpha > 0$) and infinite-range interactions ($\alpha = 0$). For the finite-range case, we work in the computational basis and explicitly represent the full Hilbert space, whose dimension scales with system size as $2^N$. 
As a result, optimization routines, which require multiple independently initialized trials to achieve convergence, become increasingly costly as the system size grows.
We therefore limit our simulations to $N = 8$, beyond which the simulations, and hence the time to reach reasonable convergence for an optimization method, become prohibitively slow. 

\Rev{In contrast, for infinite-range interactions, we use the collective Dicke states ($\ket{S,m}$) as the working basis spanning a Hilbert space that scales as $\mathcal{O}(N^2)$~\cite{PermInvariance}. Under these dynamics, axis-dependent local dephasing preserves permutational symmetry and does not generate coherences between Dicke states across different $S$ sectors. Therefore, restricting dynamics to the Dicke-basis representation reduces computational complexity and enables simulations for larger values of $N$. Further details on exploiting permutational symmetry in open quantum system dynamics can be found in Ref.~\cite{PermInvariance}.}

Moreover, the permutational invariance approach considerably aids in circumventing spectral degeneracies that arise when infinite-range interactions are implemented in a full uncoupled basis. 
Avoiding spectral degeneracies is
critical for our approach based on first-order optimization
methods; computation of the gradient of Eq.~(5) with respect to $\mathbf{x}^{(n)}$ necessitates computation of the gradients
of eigenvalues and eigenvectors with respect to $\hat{\rho}(\mathbf{x}^{(n)})$.
Elementary arguments~\cite{narayanan2024challengesdifferentiablequantumdynamics} demonstrate that the latter
gradient generally does not exist when an eigenvalue of
multiplicity greater than one is present. \Rev{We note that such eigenvalues were observed almost everywhere when working in an uncoupled basis but were not empirically observed upon exploiting permutational invariance in our infinite-range experiments.}

To support gradient-based optimization, our computation of $\hat{\rho}(\mathbf{x}^{(n)})$ and $F_Q[\hat{\rho}(\mathbf{x}^{(n)}); \hat{G}]$ must be differentiable. 
Our approach based on Ref.~\cite{narayanan2024challengesdifferentiablequantumdynamics} uses JAX~\cite{jax2018github}, a Python library that supports automatic differentiation. 
The JAX-based library Diffrax~\cite{kidger2021on} was employed to solve Eq.~\eqref{eqn:master-eqn}. Because $F_Q[\hat{\rho}(\mathbf{x}^{(n)}); \hat{G}]$ is a function of eigenvalues and eigenvectors of $\hat{\rho}(\mathbf{x}^{(n)})$, the eigenvalue decomposition operation must be differentiated. JAX computes the derivatives for {\tt jax.numpy.linalg.eigh} for the case where the input has distinct eigenvalues.
We found in practice, however, that near eigenvalue crossings, numerical instabilities result. 
Therefore, we implemented a technique for computing derivatives of eigenvalues and eigenvectors (see Ref.~\cite{chu1990multiple}, Section 4.2) to perform this derivation.

Combining these observations with elementary calculus (chain rule, product rule, and quotient rule), we may compute gradients of $F_Q[\hat{\rho}(\mathbf{x}^{(n)}); \hat{G}]$ in Eq.~\eqref{eqn:qfi} with respect to $\mathbf{x}^{(n)}$ for simulations on the computational basis or in a permutationally invariant setting of computing $\hat{\rho}(\mathbf{x}^{(n)})$ (respectively corresponding to the {\tt spin\_models} and {\tt PermSolver} APIs of Ref.~\cite{qfiopt}).
Technically, eigenvalue crossings in $\hat{\rho}(\mathbf{x}^{(n)})$ imply that $F_Q[\hat{\rho}(\mathbf{x}^{(n)}); \hat{G}]$ is not everywhere differentiable; but in our tests, evaluating a set of circuit parameters at which $F_Q[\hat{\rho}(\mathbf{x}^{(n)}); \hat{G}]$ is nondifferentiable (i.e., parameters $\mathbf{x}^{(n)}$ at which any two eigenvalues of $\hat\rho(\mathbf{x}^{(n)})$ are \emph{exactly} equal) occurs with an empirical probability of zero. 
With these gradients in hand, we employ the implementation of LBFGS-B, a quasi-Newton optimization method that respects bound constraints~\cite{lbfgsb}; the method is wrapped and distributed in SciPy~\cite{scipy}. 

We constrain the parameter set $\mathbf{x}^{(n)}$ to lie within lower ($\ell$) and upper ($u$) bounds in order to avoid any cyclicality that may be encountered across the rotational and entangling VQC stages. 
By specifically tailoring the VQC to forward state-propagation, the rotational stages are limited to timescales within $\theta_x^{k}\in [0, \pi]$. The entangling stages are similarly constrained to $\theta_{\mathrm{I}}^{k}\in [0, \pi]$, motivated by the corresponding closed-system dynamics ($\gamma/\chi = 0$), where one-axis-twisting-like evolution is periodic and exhibits state revivals. Furthermore, without loss of generality, we restrict $\varphi_1$ to $[0, 2\pi]$. Under these constraints, we search for the optimal parameter set $\mathbf{x}_{\mathrm{opt}}^{(n)}$ that maximizes the QFI; that is, we define $\mathbf{x}_{\mathrm{opt}}^{(n)}$ as the solution to the maximization problem:

\begin{equation}
    \label{eqn:opt_prob}
    \displaystyle\max_{\mathbf{x}^{(n)}} F_Q[\mathbf{x}^{(n)};\hat{G}] \quad \text{s.t.} \quad \ell \leq \mathbf{x}^{(n)} \leq u.
\end{equation}

The function landscape of $F_Q[\hat{\rho}(\mathbf{x}^{(n)}); \hat{G}]$ appears generally multimodal, meaning that any optimization method intended for local search, such as LBFGS-B, is prone to converge to suboptimal local maxima. 
However, our practical experience indicates that the landscape of $F_Q[\hat{\rho}(\mathbf{x}^{(n)}); \hat{G}]$ is not pathologically multimodal, either. 
Thus, because we are provided with natural bound constraints in Eq.~\eqref{eqn:opt_prob}, we warmstart LBFGS-B with an initial point determined by Bayesian optimization with a relatively small budget $(10 (2n + 3))$ of function evaluations. 
Bayesian optimization is a well-studied (see, e.g.,~\cite{shahriari2015taking, frazier2018bayesian} for a relevant survey and tutorial) zeroth-order method for bound-constrained global optimization that employs Gaussian processes iteratively to (i) update beliefs on a functional form for an objective ($F_Q[\hat{\rho}(\mathbf{x}^{(n)}); \hat{G}]$) landscape and (ii) query promising candidate parameters ($\mathbf{x}^{(n)}$) according to some acquisition function derived from the posterior belief. 
We employed a Python implementation of Bayesian optimization~\cite{bayesianoptimization} and found that the default hyperparameters worked well for our purposes. 

We remark again that we deliberately chose to use Bayesian optimization only for identifying a good initial point for LBFGS-B, as opposed to using Bayesian optimization for the full optimization.
We did so because Bayesian optimization cannot readily employ derivatives. 
Therefore, although Bayesian optimization is useful for performing a disciplined global exploration and often identifies a good approximation to a global maximum within a budget, it is relatively inefficient (compared with LBFGS-B) at ``fine-tuning'' the approximation into a more precise solution. 

\Rev{Because the Bayesian-optimization phase searches globally across the feasible region defined by the bound constraints of Eq.~(7), no hand-tuned or physically motivated initial guess for $\mathbf{x}^{(n)}$ is required; any point within the feasible box is, in principle, an equally valid starting point for the warmstart search. LBFGS-B was executed with a termination tolerance of $10^{-2}$, setting both the gradient- and function-value convergence criteria, or otherwise default iteration settings, whichever is reached first. To further mitigate the risk of converging to a suboptimal local maximum of the multimodal landscape described above, for each fixed decoherence rate, system size, entangling Hamiltonian, and layer number $n$ we additionally reran the full Bayesian-optimization warmstart followed by LBFGS-B refinement with independent random seeds in regions exhibiting competing locally optimized solutions, retaining the trial attaining the largest QFI; the number of such additional reruns was varied by region rather than fixed.}

\section{Optimal state preparation}\label{sec:results}
We now present results from our numerical optimization using the Ising and FTAT Hamiltonians given in Eqs.~(\ref{eqn:ising-ham}) and~(\ref{eqn:tat-ham}). In line with Ref.~\cite{qfi_opt_manuscript1} we find that increasing the decoherence strength leads to a monotonic reduction of the QFI toward the SQL; this is accompanied by a qualitative transition in the nature of the generated optimal states. In the following, we categorize the states generated by our VQC into three broad classes: (i) cat-like, (ii) squeezed-like, and (iii) uncorrelated states~\cite{qfi_opt_manuscript1}. In Sec.~\ref{sec:diagnostic-tools}, we provide a set of diagnostic tools that will aid in identifying which of the three classes generated states correspond to. Then, in Sec.~\ref{sec:inf-range-interactions}, we look at results generated with infinite-range interactions, i.e., with $\alpha = 0$.
Finally, in Sec.~\ref{sec:finite-range-interactions} we extend the analysis to the finite-range interaction regime ($\alpha = 3$) for the Ising and FTAT models as a direct comparison of VQC performance across interaction ranges. In what follows, we show that layering can offer two benefits: \Rev{it can either reduce the cumulative entangling evolution required by the optimized protocol or increase the attainable QFI.} In most cases, we observe that only one of these benefits can be harnessed at a given time. Notably, OAT and a subset of the $\alpha = 3$ Ising regime are exceptions, where layering can deliver both benefits at once.

\subsection{Diagnostic Tools}\label{sec:diagnostic-tools}
We now introduce a set of diagnostics used to characterize the states generated by the VQC. Broadly, these states fall into three classes. First, cat-like states are composed of macroscopic superpositions. Second, squeezed-like states exhibit reduced fluctuations along a quadrature orthogonal to a common polarization axis. Finally, uncorrelated states are collectively polarized along a common axis and contain no useful inter-particle correlations~\cite{spin-squeezing-ueda, qfi_opt_manuscript1}.

\subsubsection{State fidelities}\label{sec:diagnostic-tools-fidelity}
We define the general fidelity between a pure target state $\ket{\Psi}$ and the $n$-layer VQC output state $\hat{\rho}(\mathbf{x}_{\mathrm{opt}}^{(n)})$ as
\begin{equation}\label{eqn:general-fidelity}
    \mathcal{F}^{(n)}_{\Psi} = \bra{\Psi}\hat{\rho}(\mathbf{x}_{\mathrm{opt}}^{(n)})\ket{\Psi}.
\end{equation}
In our results, we compute target-state fidelities with specific pure states. For example, we compute GHZ fidelities ($\mathcal{F}_{\mathrm{GHZ}}^{(n)}$) by setting $\ket{\Psi} = \ket{\mathrm{GHZ}, \Phi}$ [see Eq.~(\ref{eqn:ghz-state})]. The fidelity is maximized by optimizing over the phase $\Phi$. In this way, we classify states with appreciable GHZ overlaps as cat-like, i.e., $\mathcal{F}_{\mathrm{GHZ}}^{(n)}\geq0.5$. 

In the text we delineate the $1$-layer cat-like regime by decoherence $\gamma_1$ which marks the threshold at which $\mathcal{F}_{\mathrm{GHZ}}^{(1)}$ drops below $0.5$. In Figs.~\ref{fig:OAT_v_gamma} and~\ref{fig:TAT_v_gamma}, $\gamma_1$ is indicated by a vertical dashed line. Moreover, $\gamma_1$ sets a baseline cat-like regime to compare against when increasing the E--R layer count. 

\subsubsection{Cumulative quantities}
In the text we also present cumulative quantities for an $n$-layer VQC, namely the cumulative entangling strength (CES) and cumulative rotation angle (CRA)
\begin{eqnarray}\label{eqn:cum-ent-strength}
    \Theta_{\mathrm{I}}^{(n)} &=& \sum_{k=1}^n \theta_{\mathrm{I}}^k,\\\label{eqn:cum-rot-angle}
    \Theta_x^{(n)} &=& \sum_{k=1}^{n}{\theta_x^k}.
\end{eqnarray}
Equations~(\ref{eqn:cum-ent-strength}) and~(\ref{eqn:cum-rot-angle}) provide a compact characterization of multi-layer VQC implementations, whose number of parameters increases as $2n + 3$ with layer number. 

In the text we use the CES to delineate the uncorrelated regime at the decoherence $\gamma_2$ where $\Theta_{\mathrm{I}}^{(1)} = 0$. At this threshold, $\Theta_{\mathrm{I}}^{(n)} = 0$ for $n = 1$ and $n = 3$ E--R layers indicate that the VQC strategy prescribes no entanglement generation. In Figs.~\ref{fig:OAT_v_gamma}-\ref{fig:local_TAT_v_gamma}, the onset of the uncorrelated regime is indicated by a vertical dashed line situated in the limit of strong decoherence (i.e., $\gamma/\chi \gg1$). In the cases presented, $\gamma_2$ is the same for both $1$- and $3$-layer VQCs. As no entanglement is generated within the uncorrelated regime, the states remain separable product states throughout the VQC sequence and are oriented to the Bloch equator by the VQC rotations.

\subsubsection{Squeezing}
As another diagnostic, we compute the Wineland squeezing parameter~\cite{wineland-spin-squeezing}, 
\begin{equation}\label{eqn:squeezing}
    \xi^2_s = N\min_{\mathbf{n}_\perp} \frac{(\Delta \hat{S}_{\mathbf{n}_\perp})^2}{|\braket{\hat{\mathbf{S}}}|^2}, 
\end{equation}
where $\hat{\mathbf{S}} = (\hat{S}_x, \hat{S}_y, \hat{S}_z)$ is the spin vector operator, $(\Delta\hat{S}_{\mathbf{v}})^2 = \braket{(\hat{\mathbf{S}}\cdot\mathbf{v})^2} - \braket{\hat{\mathbf{S}}\cdot\mathbf{v}}^2$ is the spin variance along axis $\mathbf{v}$, and the minimization is performed over unit vectors $\mathbf{n}_\perp$ orthogonal to the mean spin vector $\braket{\hat{\mathbf{S}}}$. In the text, we present the scaled squeezing parameter $0\leq (N\xi^2_s)^{-1}\leq F_Q/N^2$, where values above the SQL threshold [$(N\xi^2_s)^{-1}\geq1/N$] indicate metrologically useful squeezing.

\subsubsection{Classical Fisher Information \& State Collectivity}\label{sec:diagnostic-tools-cfi-collectivity}
We compute the state's classical Fisher information (CFI)~\cite{RevModPhys.90.035005, caves_1994_qfi},
% \begin{multline} \label{eqn:cfi}
% \Rev{F_C(\theta_0) = \lim_{\delta \theta\to0} \frac{4}{\delta \theta^2} \sum_{m_z} \left(\sqrt{P(m_z\vert\theta_0 + \delta \theta)} - \sqrt{P(m_z\vert\theta_0)}\right)^2,}
% \end{multline}
\begin{equation}\label{eqn:cfi}
    \Rev{F_C(\theta_0) = \lim_{\delta \theta\to0} \frac{4}{\delta \theta^2} \sum_{m_z} \left(\sqrt{P(m_z\vert\theta_0 + \delta \theta)} - \sqrt{P(m_z\vert\theta_0)}\right)^2,}
\end{equation}
\Rev{where $P(m_z\vert\theta_0 + \delta\theta)$ is the conditional probability for the collective spin projection $m_z$ along $\hat{z}$ after the state is rotated by a small angle $\delta \theta$ from the reference $\theta_0$\footnote{\Rev{Equation~(\ref{eqn:cfi}) follows from the standard expression $F_C(\theta) = \sum_x [\partial_{\theta}P(x\vert\theta)]^2/P(x\vert\theta) \equiv 4\sum_x\left[\partial_{\theta}\sqrt{P(x\vert\theta)}\right]^2$ after expressing the derivative as a finite-difference limit about $\theta_0$ and taking $x\to m_z$.}}. Without loss of generality, we set $\theta_0 =0$ in our calculations.} Bounded from above by the QFI, the CFI quantifies the extent to which readout restricted to collective observables may exploit a given state's metrological potential. 

We also compute the state collectivity $\langle \hat{S}^2\rangle/S_{\mathrm{max}}$ with $\hat{S}^2 = \hat{\mathbf{S}}\cdot\hat{\mathbf{S}}$. This quantity tracks how much of the state remains in the fully collective Dicke sector with $S_{\mathrm{max}} = (N/2)(N/2 + 1)$.

\subsection{Infinite-range interactions, $\alpha = 0$}\label{sec:inf-range-interactions}

\begin{figure}
    \centering
    \includegraphics[width=1.05\linewidth]{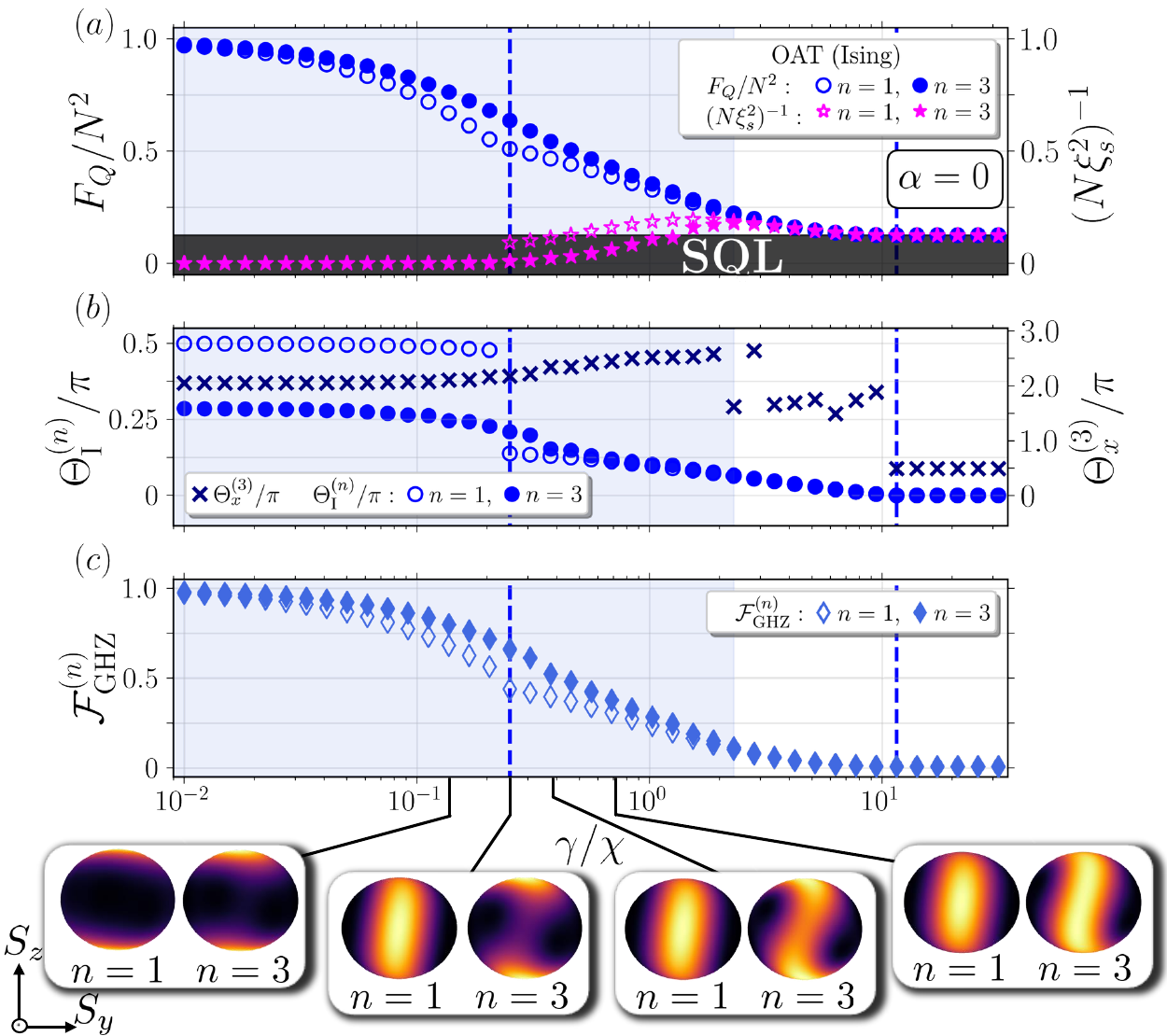}
    \caption{A set of state diagnostic quantities and VQC parameters with an OAT (Ising with $\alpha = 0$) Hamiltonian presented as functions of decoherence strength, $\gamma/\chi$. Open (filled) markers for $n = 1(3)$ E--R layers. (a) Optimal QFI ($F_Q/N^2$, circles) and rescaled squeezing parameter [$(N\xi^2_s)^{-1}$, stars].
    (b) The cumulative VQC interaction strengths ($\Theta_{\mathrm{I}}^{(n)}/\pi$, circles) and $3$-layer $x$-rotations ($\Theta^{(3)}_x/\pi$, $\times$). (c) State fidelity with a GHZ state $(\mathcal{F}^{(n)}_{\mathrm{GHZ}}$, diamonds). Vertical dashed lines in all panels indicate $\gamma_1$ and $\gamma_2$, and the shaded region highlights the MIR. 
    Within the MIR, the $3$-layer protocol exhibits more robust QFI and GHZ fidelities than does the $1$-layer counterpart. Moreover, signatures of $\Theta_{\mathrm{I}}^{(3)}< \Theta_{\mathrm{I}}^{(1)}$ \Rev{indicate a reduction in cumulative entangling evolution for the optimized $3$-layer protocol.} (Below) Comparisons of Husimi-like distributions for generated states with a $1$- and $3$-layer VQC architecture with brighter regions corresponding to higher probability densities.}
    \label{fig:OAT_v_gamma}
\end{figure}

Having outlined a set of diagnostic tools, we now analyze the states generated with the Ising and FTAT Hamiltonians of Eqs.~(\ref{eqn:ising-ham}) and~(\ref{eqn:tat-ham}) with $\alpha = 0$. In this case, the qubit-qubit spacing dependence vanishes and the coupling satisfies $\chi_{ij} = \chi$, so that the power-law Hamiltonians reduce to the one-axis twisting (OAT; $\propto\hat{S}_z^2$) and two-axis twisting [TAT; $\propto(\hat{S}_x\hat{S}_y + \hat{S}_y\hat{S}_x)$] Hamiltonians. In the following, we present results for $N = 8$ and leverage the well-studied dynamics of OAT and TAT~\cite{spin-squeezing-ueda, tat-witkowska} for unitary evolution (i.e., with $\gamma/\chi=0$) to develop intuition for our results in the presence of decoherence.

\subsubsection{One-axis twisting}\label{sec:OAT}

Results for an E--R-layered OAT scheme are presented in Fig.~\ref{fig:OAT_v_gamma} as a function of the scaled decoherence strength $\gamma/\chi$. Panel~\ref{fig:OAT_v_gamma}(a) shows the optimal QFI for $1$-layer ($n = 1$, open circles) and 3-layer ($n = 3$, filled circles) OAT. 
The QFI of each case is scaled by the HL ($F_Q/N^2$). 
For weak decoherence (i.e., $\gamma/\chi \ll 1$), the $1$-layer OAT VQC already generates states with QFIs near the fundamental HL. The QFI monotonically decreases with $\gamma/\chi$ and approaches the SQL (marked by the horizontal shaded region) for $\gamma/\chi \gg 1$. To supplement the QFI, Figure~\ref{fig:OAT_v_gamma}(b) shows the CES [$\Theta_{\mathrm{I}}^{(n)}$; see Eq.~(\ref{eqn:cum-ent-strength})] for $n$ layers. As a primary feature, the $1$-layer CES (open circles) decreases abruptly at $\gamma/\chi \approx 0.25$ and coincides with the slight kink in the QFI. 

Using the fidelity criterion outlined in Sec.~\ref{sec:diagnostic-tools-fidelity}, the OAT cat-like regime is given by $\gamma/\chi< \gamma_1/\chi \approx0.25$ [see vertical dashed line in Figs.~\ref{fig:OAT_v_gamma}(a)-\ref{fig:OAT_v_gamma}(c)], which also coincides with the abrupt transition in $\Theta_{\mathrm{I}}^{(1)}$. Within this regime, $\Theta_{\mathrm{I}}^{(1)} \approx \pi/2$ (and the full parameter set $\mathbf{x}_{\mathrm{opt}}^{(n)}$) aligns with known strategies for the generation of GHZ states~\cite{Foss-FeigM2013Dqco-ising-cat-states, qfi_opt_manuscript1}. The GHZ fidelity further clarifies the nature of the generated states [$\mathcal{F}_{\mathrm{GHZ}}^{(1)}$; see open diamonds in panel (c)]. The fidelity begins at $\mathcal{F}^{(1)}_{\mathrm{GHZ}}\approx 1.0$ for $\gamma/\chi= 10^{-2}$ and retains appreciable GHZ overlaps (i.e., $\mathcal{F}^{(1)}_{\mathrm{GHZ}} \geq 0.5$) for $\gamma \lesssim \gamma_1$.

For decoherence $\gamma > \gamma_1$, both the GHZ fidelity and $\Theta_{\mathrm{I}}^{(1)}$ decrease monotonically before vanishing at the onset of the uncorrelated regime at $\gamma_2 / \chi \approx 11.55$ [see vertical dashed line in panels (a)-(c)]. We therefore focus on the intermediate window $\gamma_1 < \gamma < \gamma_2$, which we designate as the squeezed-like regime because the optimized preparation strategies $\mathbf{x}_{\mathrm{opt}}^{(n)}$ reproduce OAT's well-known generation of squeezed states~\cite{spin-squeezing-ueda}. 

Within the cat-like regime, the scaled squeezing parameter $(N\xi_s^2)^{-1} = 0$ [open stars; see panel~\ref{fig:OAT_v_gamma}(a)]. These values are consistent with spin-vector depolarization (i.e., $\vert\braket{\mathbf{S}}\vert \approx 0$) and are supported by appreciable GHZ state fidelities. At $\gamma_1$, $\mathcal{F}^{(1)}_{\mathrm{GHZ}}< 0.5$ while $(N\xi_s^2)^{-1}$ increases abruptly. Therefore, we interpret $\gamma_1$ as the crossover between cat-like and squeezed-like regimes. Within the squeezed-like regime, $(N\xi_s^2)^{-1}$ rises above the SQL before decreasing and settling at the SQL for $\gamma \geq \gamma_2$. Accordingly, values of $(N\xi_s^2)^{-1} > 1/N$ correspond to squeezed states that are strongly polarized ($\vert\braket{\hat{\mathbf{S}}}\vert \approx N/2$) and have reduced transverse fluctuations, $(\Delta \hat{S}_{\mathbf{n}_{\perp}})^2 < N/4$, which together enhance Ramsey spectroscopy~\cite{spin-squeezing-ueda, wineland-spin-squeezing, quantum-spin-squeezing}.

We next consider whether increasing the number of E--R layers can improve the performance across the decoherences considered. To this end, we increase the VQC E--R layer count to $n = 3$ for the same OAT Hamiltonian. To compare with the $1$-layer OAT case (open markers), we show the corresponding $3$-layer results using filled markers in panels~\ref{fig:OAT_v_gamma}(a)--\ref{fig:OAT_v_gamma}(c). We find that increasing the E--R layer count yields a more robust QFI within the metrological improvement regime (MIR) $10^{-2}\lesssim \gamma/\chi \lesssim 2.3$ [see shaded region in Fig.~\ref{fig:OAT_v_gamma}]. Note that outside of the MIR the QFI and supplemental quantities are virtually identical to the $1$-layer case.

To better understand the QFI gain of layered OAT across the MIR we turn to the $3$-layer CES ($\Theta_{\mathrm{I}}^{(3)}/\pi$). In Fig.~\ref{fig:OAT_v_gamma}(b), the $3$-layer CES begins at $\Theta_{\mathrm{I}}^{(3)}/\pi \approx 0.29 $ for $\gamma /\chi =10^{-2}$ and remains smaller than $\Theta_{\mathrm{I}}^{(1)}/\pi$ throughout the cat-like regime (i.e., $\gamma < \gamma_1$). \Rev{Accordingly, the layered strategy requires less cumulative entangling duration while reaching states of comparable or better QFI.}

More specifically, the intermediate $x$-rotations either help to reduce cumulative entangling duration over part of the MIR or sustain QFI enhancement when the $1$- and $3$-layer CESs are comparable ($\Theta_{\mathrm{I}}^{(3)}\approx\Theta_{\mathrm{I}}^{(1)}$). The $3$-layer CRA [$\Theta_x^{(3)}$; see Eq.~(\ref{eqn:cum-rot-angle})] is shown in panel~\ref{fig:OAT_v_gamma}(b). Within the MIR, the growth of $\Theta_x^{(3)}$ is accompanied by a decrease of $\Theta_{\mathrm{I}}^{(3)}$ well into the squeezed-like regime.

As a consequence of layering, the cat-like regime becomes more robust across the MIR. Specifically, we observe larger $3$-layer GHZ fidelities at larger $\gamma/\chi$ relative to the $1$-layer case [see closed diamonds in panel~\ref{fig:OAT_v_gamma}(c)]. 
Layering thus extends the cat-like regime beyond $\gamma_1$, as $\mathcal{F}_{\mathrm{GHZ}}^{(3)}$ remains above $0.5$ up to $\gamma/\chi \approx 0.375$. Within the MIR the separation between $\mathcal{F}_{\mathrm{GHZ}}^{(3)}$ and $\mathcal{F}_{\mathrm{GHZ}}^{(1)}$ is comparable to that seen between the $1$- and $3$-layer QFIs. 

\begin{figure}[b]
    \centering
    \includegraphics[width=1.0\linewidth]{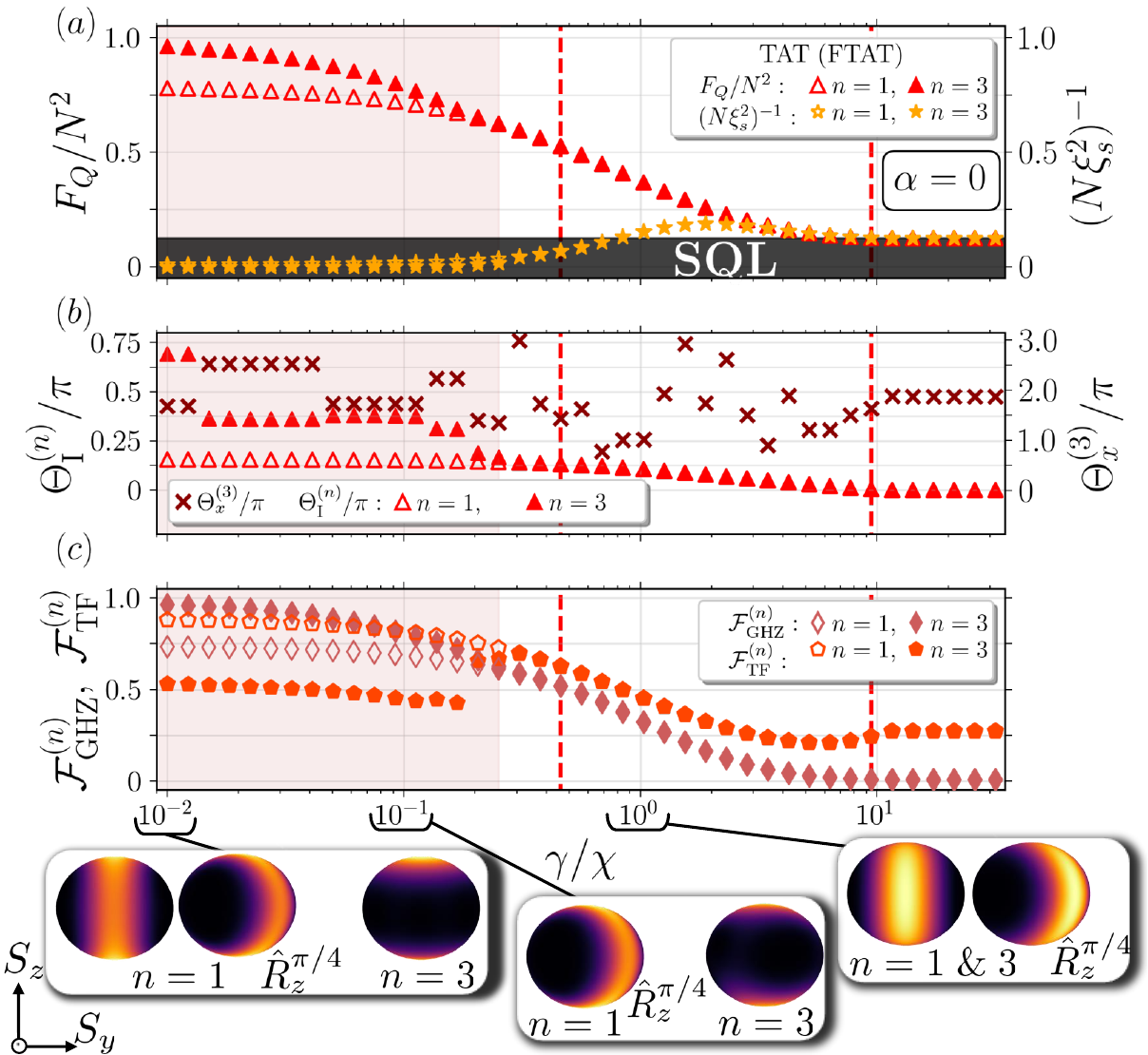}
    \caption{A set of state diagnostic quantities and VQC parameters with a TAT (FTAT with $\alpha = 0$) Hamiltonian presented as functions of decoherence strength, $\gamma/\chi$. Open (filled) markers for $n = 1(3)$ E--R layers. (a) Optimal QFI ($F_Q/N^2$, triangles) and rescaled squeezing parameter [$(N\xi^2_s)^{-1}$, stars.
    (b) Cumulative VQC interactions ($\Theta_{\mathrm{I}}^{(n)}/\pi$, triangles) and $3$-layer $x$-rotations ($\Theta^{(3)}_x/\pi$, $\times$). (c) State fidelity with the GHZ state $(\mathcal{F}^{(n)}_{\mathrm{GHZ}}$, diamonds) and TF state $(\mathcal{F}^{(n)}_{\mathrm{TF}}$, pentagons). Vertical dashed lines in all panels indicate $\gamma_1$ and $\gamma_2$, and the shaded region highlights the MIR. Within the MIR, the decrease in $\mathcal{F}_{\mathrm{TF}}^{(n)}$ and increase in $\mathcal{F}_{\mathrm{GHZ}}^{(n)}$ and in the QFI upon increasing the layer count to $n = 3$ suggest the generation of $\mathrm{GHZ}$ and cat-like states with TAT. (Below) Husimi-like distribution comparisons for generated states with a $1$- and $3$-layer VQC architecture with brighter regions corresponding to higher probability densities. States with a $\hat{R}_z^{\pi/4}$ designation depict the state rotated by an angle of $\pi/4$ about $z$.}
    \label{fig:TAT_v_gamma}
\end{figure}

The shift toward cat-like character is further reflected by a reduction in $3$-layer $(N\xi_s^2)^{-1}$ (filled stars). To this end, we depict a handful of Husimi-like distributions (see Appendix~\ref{sec:husimi} for definitions and further detail) below Fig.~\ref{fig:OAT_v_gamma} (and in Figs.~\ref{fig:TAT_v_gamma}--\ref{fig:local_TAT_v_gamma}) to qualitatively show the states generated with the $1$- and $3$-layer VQCs. Notably, the $3$-layer VQC sustains cat-like features for greater decoherence relative to the $1$-layer case.

As a whole, layering yields two concrete benefits for OAT: (i) it increases the QFI robustness across the MIR, and (ii) it achieves higher or comparable QFI with a reduced or comparable CES $\Theta_{\mathrm{I}}^{(3)} \lesssim \Theta_{\mathrm{I}}^{(1)}$. \Rev{The latter benefit reduces the cumulative duration over which decoherence acts during the entangling stages.} Together, these features extend the cat-like regime.

\subsubsection{Two-axis twisting}\label{sec:TAT}

Having identified benefits of layering in the OAT case, we now apply a similar analysis to VQCs implemented with a TAT Hamiltonian (FTAT with $\alpha = 0$). Figure~\ref{fig:TAT_v_gamma} summarizes the TAT performance as a function of $\gamma/\chi$. In panel~\ref{fig:TAT_v_gamma}(a), the scaled QFI ($F_Q/N^2$) is shown for $n = 1$ (open triangles) and $n = 3$ (filled triangles) VQC architectures. The $1$-layer VQC acquires a QFI of $F_Q/N^2 \approx 0.78$ at $\gamma/\chi = 10^{-2}$, while increasing the layer count to $n = 3$ yields a substantial enhancement to $F_Q/N^2\approx 0.96$. The advantage given by layering diminishes at stronger decoherence. Specifically, for $\gamma/\chi \gtrsim 0.25$, the $n = 1$ and $n = 3$ QFIs converge and layering no longer yields a metrological advantage. The supplemental diagnostic quantities given in panels~\ref{fig:TAT_v_gamma}(a)--\ref{fig:TAT_v_gamma}(c) similarly converge. We therefore focus on the corresponding TAT MIR $10^{-2} \leq \gamma/\chi \leq 0.25$ (located within the cat-like regime, $\gamma_1 \lesssim 0.25$) where layering provides a clear QFI improvement.

In Fig.~\ref{fig:TAT_v_gamma}(b) we present $n$-layer CES ($\Theta_{\mathrm{I}}^{(n)}/\pi$) and CRA ($\Theta_{x}^{(n)}/\pi$; $\times$). For the $1$-layer case, the CES is approximately constant at $\Theta_{\mathrm{I}}^{(1)}/\pi \approx 0.16$ before monotonically decreasing to zero at $\gamma_2/\chi \approx 9.44$ (see vertical dashed line). On the other hand, the $3$-layer CES is relatively greater than the $1$-layer counterpart. Throughout the MIR, both $\Theta_{\mathrm{I}}^{(3)}$ and $\Theta_x^{(3)}$ exhibit discontinuities and ``plateaus''. Each plateau maintains approximately identical VQC parameters for substantial extents of $\gamma/\chi$ (see Appendix~\ref{sec:state-prep-strategies} for full state preparation strategies). 

We further assess the generated states by comparing the VQC states to the twin-Fock (TF) state. We use the TF state as a relevant benchmark because TF-like character is known to arise in native-TAT dynamics~\cite{tat-witkowska, PhysRevA.90.062132}. Explicitly, the TF state
\begin{equation} \label{eqn:tf-state}
    \ket{\mathrm{TF}} = \binom{N}{N/2}^{-1/2}\sum_{\substack{
    x \in \{0, 1\}^N \\
    |x| = N/2}}{\ket{x}}
\end{equation}
is composed of a balanced superposition of computational basis states having half of its qubits in $\ket{\uparrow} = \ket{1}$. For example, for $N = 4$ the bit string $x=1010$ contains half of its entries ($\vert x \vert = 2$) set to $1$. More specifically, the TF state is a Dicke state with $S = N/2$ and $m = 0$, so that $\hat{S}^2\ket{N/2, 0} = (N/2)(N/2+1)\ket{N/2, 0}$~\cite{twin-fock, tat-witkowska, PhysRevA.90.062132}.

Comparing the TF and GHZ fidelities reveals that layering shifts the preparation of states from TF-like toward more GHZ-like character within the MIR. In Fig.~\ref{fig:TAT_v_gamma}(c), the $1$-layer TF fidelity remains above the GHZ fidelity throughout the MIR ($\mathcal{F}_{\mathrm{TF}}^{(1)} > \mathcal{F}_{\mathrm{GHZ}}^{(1)}$). We note that $\mathcal{F}^{(1)}_{\mathrm{TF}}\approx 0.88$ and the corresponding QFI at $\gamma/\chi = 10^{-2}$ coincide with values reported in Ref.~\cite{tat-witkowska}, and attribute deviations to decoherence and finite-size effects from our treatment with $N = 8$. Increasing the E--R layer number to $n = 3$ results in a trade-off in the GHZ and TF fidelities (i.e., $\mathcal{F}_{\mathrm{GHZ}}^{(3)} > \mathcal{F}_{\mathrm{TF}}^{(3)}$). Specifically, the increase in GHZ fidelity and drop in $(N\xi^2_s)^{-1}$ indicate stronger cat-like character. That is, states acquire stronger GHZ character (e.g., $\mathcal{F}_{\mathrm{GHZ}}^{(3)}\approx0.97$ at $\gamma/\chi = 10^{-2}$) and the scaled squeezing parameter vanishes, $(N\xi^2_s)^{-1} = 0$.

As a whole, layering extends TAT's state space and enables the generation of GHZ-like states~\cite{gietka2021simulating}. While layering prescribes larger CESs for TAT ($\Theta_{\mathrm{I}}^{(3)} > \Theta_{\mathrm{I}}^{(1)}$), this difference may be understood as a finite-size effect that weakens with increasing $N$. This is because entanglement generation within the VQC is governed by TAT dynamics, which are known to accelerate with increasing system size~\cite{PhysRevA.90.062132, tat-witkowska}. We further discuss this feature in Sec.~\ref{sec:system-size}, where we find decreasing CESs and sustained cat-like state generation with increasing system size.

Across the OAT and TAT sections, our results show that layering promotes the generation of cat-like states. \Rev{For OAT, the E--R VQC architecture extends cat-like robustness to greater decoherence while, over part of this regime, requiring reduced cumulative entangling evolution ($\Theta_{\mathrm{I}}^{(3)}<\Theta_{\mathrm{I}}^{(1)}$).} On the other hand, layering enables the generation of GHZ and GHZ-like states in TAT by expanding the accessible state space~\cite{gietka2021simulating, tat-witkowska} (see Appendix~\ref{sec:driven_hams} for further detail). 

\subsection{Finite-range interactions, $\alpha = 3$}\label{sec:finite-range-interactions}

In this section we extend the analysis to the finite-range interaction regime ($\alpha = 3$) for the Ising and FTAT models as a direct comparison with the collective OAT and TAT cases. Although increasing $\alpha$ reduces the effective interaction range, finite-range implementations exhibit behavior similar to the collective counterparts. In particular, we find that layering significantly improves the QFI of Ising interactions. Specifically, native Ising dynamics ($n = 1$) are constrained to generating squeezed-like states. Increasing the layer count shifts preparation strategies toward the generation of cat-like states. In contrast, increasing the E--R layer count for FTAT increases the QFI only marginally, with the noticeable gain occurring for weaker decoherence.

\subsubsection{Ising}\label{sec:ising}

\begin{figure}
    \centering
    \includegraphics[width=1.0\linewidth]{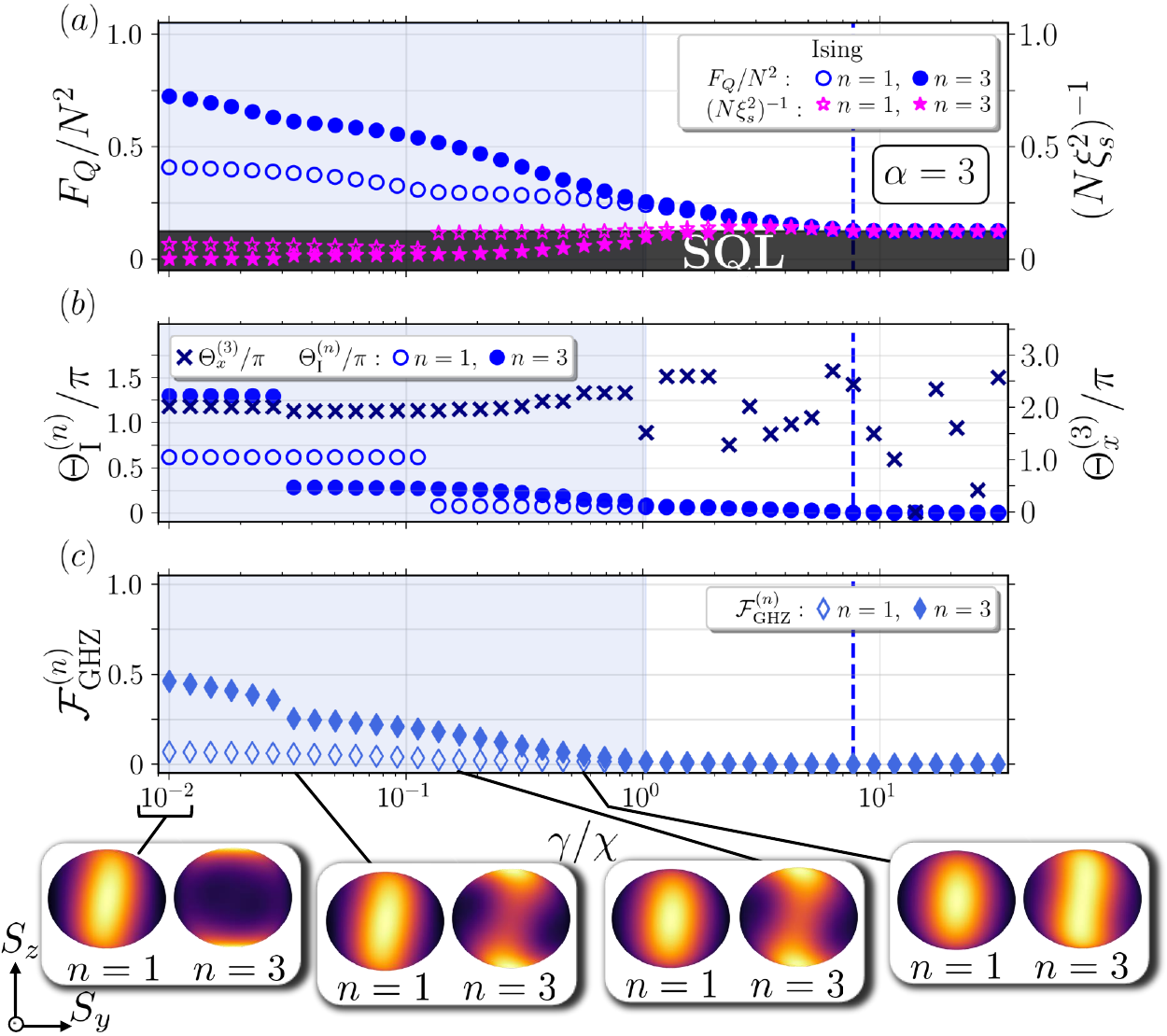}
    \caption{Diagnostic quantities and VQC parameters with an Ising Hamiltonian with $\alpha = 3$ presented as functions of decoherence strength, $\gamma/\chi$. Open (filled) markers for $n = 1(3)$ E--R layers. (a) Optimal QFI ($F_Q/N^2$, circles) and rescaled squeezing parameter [$(N\xi^2_s)^{-1}$, stars].
    (b) Cumulative VQC interaction strengths ($\Theta_{\mathrm{I}}^{(n)}/\pi$, circles) and $3$-layer $x$-rotations ($\Theta^{(3)}_x/\pi$, $\times$). (c) State fidelity with a GHZ state $(\mathcal{F}^{(n)}_{\mathrm{GHZ}}$, diamonds). Vertical dashed lines in all panels indicate $\gamma_2$, and the shaded region highlights the MIR. Notable increases in the QFI occur within the MIR upon increasing the layer count to $n = 3$. (Below) Husimi-like distribution comparisons for generated states with a $1$- and $3$-layer VQC architecture with brighter regions corresponding to higher probability densities.}
    \label{fig:ising_v_gamma}
\end{figure}

In this section we compare the results for $\alpha = 3$ Ising to the collective OAT case presented in Sec.~\ref{sec:OAT}. Relative to OAT, the shortened interaction ranges of finite-range Ising attenuate the achievable QFIs [see Fig.~\ref{fig:ising_v_gamma}(a)]. Nevertheless, layering significantly improves the achievable QFI within the Ising implementation. Specifically, increasing the layer count from $n = 1$ to $n = 3$ at $\gamma/\chi = 10^{-2}$ improves the achieved QFI from $F_Q/N^2\approx 0.41$ to $F_Q/N^2\approx 0.72$. This enhancement persists throughout the Ising MIR given by $\gamma/\chi \lesssim 1.0$ (see shaded region).

Despite having comparatively weaker QFIs than in the OAT case, Ising preparation strategies showcase similar behavior. For instance, there exist abrupt changes between CES ($\Theta_{\mathrm{I}}^{(n)}$) plateaus that align with subtle kinks in the QFI.  Moreover, \Rev{we identify regimes in which layering reduces the cumulative entangling duration.} Elsewhere, the $3$-layer CES is greater ($\Theta_{\mathrm{I}}^{(1)} \lesssim \Theta_{\mathrm{I}}^{(3)}$). In either case, layered implementations yield a greater or comparable QFI. For the $3$-layer case specifically, an increasing CRA ($\Theta_x^{(3)}$) coincides with a decreasing CES ($\Theta_{\mathrm{I}}^{(3)}$).

In contrast to the case of OAT, $1$-layer Ising implementations constrain state generation to squeezed-like states. In particular, finite scaled squeezing parameters [$(N\xi^2_s)^{-1}$, open stars] across $\gamma/\chi$ are consistent with squeezed-like character. Additionally, the $1$-layer case is devoid of a cat-like regime as the GHZ fidelity remains below the cat-like threshold (i.e., $\mathcal{F}_{\mathrm{GHZ}}^{(1)}<0.5$). Therefore, we designate the squeezed-like regime for decoherence up to the uncorrelated regime ($\gamma/\chi \lesssim \gamma_2/\chi\approx7.9$, see vertical dashed line).

Increasing the layer count to $n = 3$ expands the native Hamiltonian's accessible state space by shifting the generation strategies toward cat-like states. Most notably, layering narrowly enables the emergence of a cat-like regime for $\gamma/\chi = 10^{-2}$ as the GHZ fidelity begins to probe the cat-like threshold (i.e., $\mathcal{F}_{\mathrm{GHZ}}^{(3)}\approx 0.5$). Alongside an increase in the GHZ fidelity, sharp decreases in $(N\xi^2_s)^{-1}$ relative to the $1$-layer case further indicate state depolarization in line with cat-like character. As we will confirm in Sec.~\ref{sec:interaction-range}, increasing the layer count beyond $n = 3$ yields systematic QFI improvements that approach the fundamental HL. 

\subsubsection{FTAT}\label{sec:ftat}

\begin{figure}[b]
    \centering
    \includegraphics[width=1.0\linewidth]{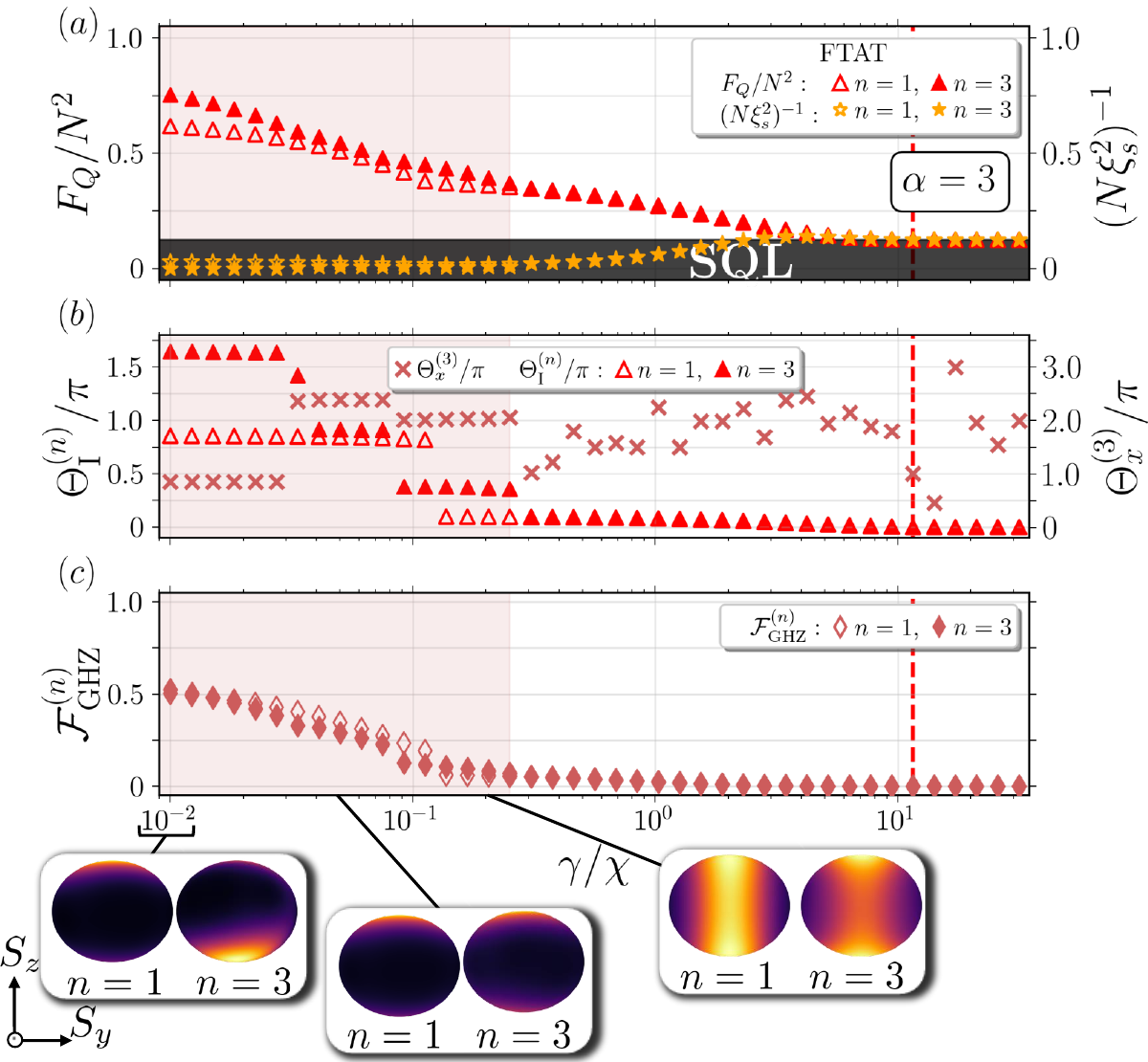}
    \caption{Diagnostic quantities and VQC parameters with an FTAT Hamiltonian with $\alpha = 3$ presented as functions of decoherence strength, $\gamma/\chi$. Open (filled) markers for $n = 1(3)$ E--R layer VQC configurations. (a) Optimal QFI ($F_Q/N^2$, triangles) and rescaled squeezing parameter [$(N\xi^2_s)^{-1}$, stars].
    (b) Cumulative VQC interaction strengths ($\Theta_{\mathrm{I}}^{(n)}/\pi$, triangles) and $3$-layer $x$-rotations ($\Theta^{(3)}_x/\pi$, $\times$). (c) Generated state fidelity with a GHZ state $(\mathcal{F}^{(n)}_{\mathrm{GHZ}}$, diamonds). Vertical dashed lines in all panels indicate $\gamma_2$, and the shaded region highlights the MIR. Marginal increases in the QFI occur within the MIR upon increasing the layer count to $n = 3$. (Below) Husimi-like distribution comparisons for generated states with a $1$- and $3$-layer VQC architecture with brighter regions corresponding to higher probability densities.}
    \label{fig:local_TAT_v_gamma}
\end{figure}

 We now turn to FTAT implementations with $\alpha = 3$ and compare them with the collective TAT case of Sec.~\ref{sec:TAT}. Relative to the Ising case of the previous section, FTAT is even less responsive to layering at $\alpha = 3$ [see Fig.~\ref{fig:local_TAT_v_gamma}(a)]. Specifically, the QFI improves modestly when increasing the layer count from $n = 1$ to $n = 3$ within the FTAT MIR ($\gamma/\chi \lesssim 0.2$, shaded region). 

While FTAT acquires limited QFI improvement, the preparation strategies are similar to the case of TAT. Specifically, the $1$- and $3$-layer state preparation strategies exhibit plateaus in $\Theta_{\mathrm{I}}^{(n)}$ and $\Theta_x^{(3)}$. Additionally, the abrupt transitions linking plateaus accompany slight kinks in the QFI. Moreover, FTAT features instances where layering requires greater CES ($\Theta_{\mathrm{I}}^{(3)} > \Theta_{\mathrm{I}}^{(1)}$) to improve the QFI. 

\begin{figure*}
    \centering
    \includegraphics[width=0.75\linewidth]{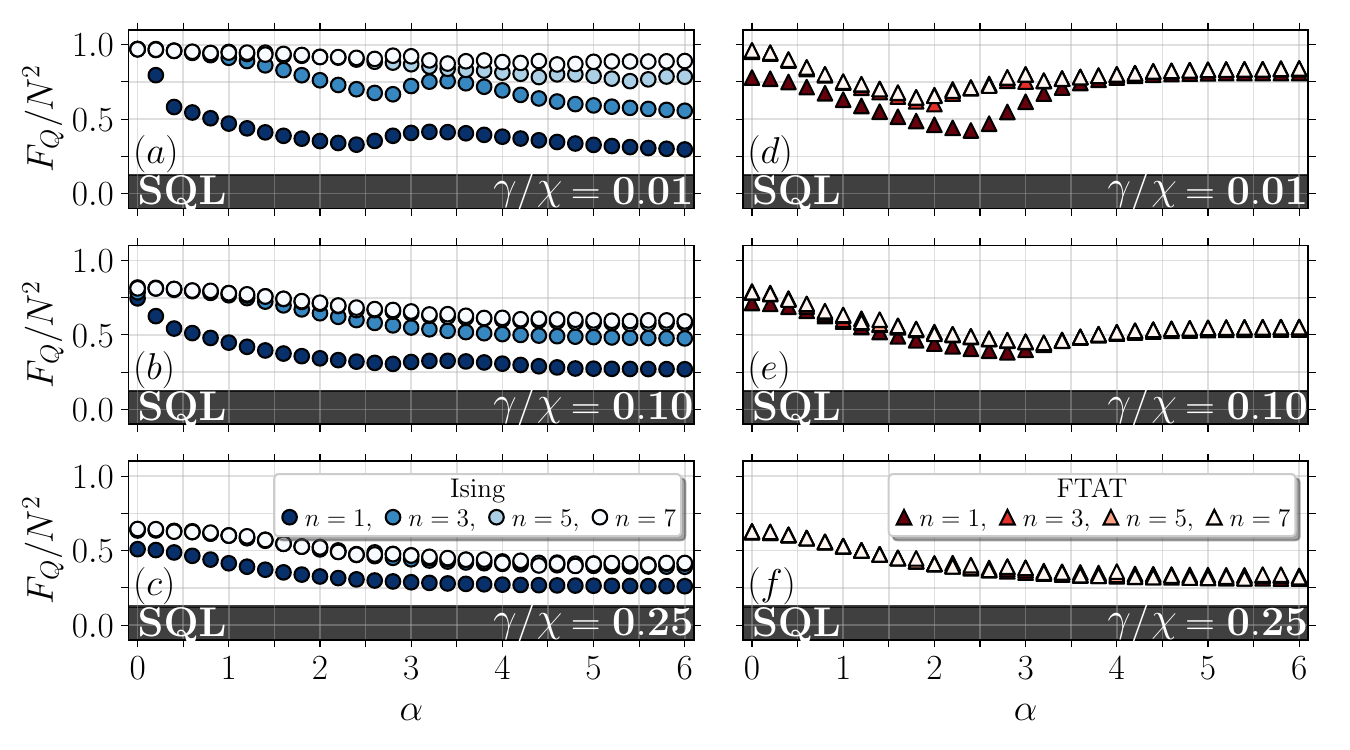}
    \caption{QFI ($F_Q/N^2$) as a function of interaction range $\alpha$ for Ising [(a)--(c), circles] and FTAT [(d)--(f), triangles] Hamiltonians. In each panel we show results for VQC implementations with $n = 1$, $3$, $5$, and $7$ layers in order of decreasing color opacity. The decoherence is held at $\gamma/\chi = 0.01$, $0.1$, and $0.25$ for the top, middle, and bottom subplots, respectively. All results computed with $N = 8$.}
    \label{fig:v-alpha}
\end{figure*}

Although the optimized strategies $\mathbf{x}_{\mathrm{opt}}^{(n)}$ are similar, FTAT yields little change in the nature of the generated states. For $\gamma/\chi = 10^{-2}$, both the $1$- and $3$-layer GHZ fidelities approach the onset of the cat-like regime (i.e., $\mathcal{F}_{\mathrm{GHZ}}^{(n)}\approx0.5$). Across the MIR, layering offers no relative GHZ state improvement. As we will show in the next section, FTAT becomes more constrained with decreasing interaction range or increasing decoherence.

\section{Interaction range}\label{sec:interaction-range}

We now analyze the role of E--R layering in the generation of states across different interaction ranges for a choice of $N = 8$. Figure~\ref{fig:v-alpha} shows the QFI ($F_Q/N^2$) for $n = 1$, $3$, $5$, and $7$ E--R layered Ising [(a)--(c)] and FTAT [(d)--(f)] VQC implementations as a function of $\alpha$, which parameterizes the decay of the interaction strength. Decoherence increases from top to bottom, with $\gamma/\chi = 0.01$, $0.1$, and $0.25$, respectively. For $\alpha = 0$ and $\alpha = 3$, the $1$- and $3$-layer Ising and FTAT QFIs coincide with the values reported in Secs.~\ref{sec:inf-range-interactions} and~\ref{sec:finite-range-interactions}.

For the Ising model, increasing the E--R layer count yields systematic QFI improvements across the decoherence strengths considered. In particular, adding layers lifts the $1$-layer Ising QFI trend across $\alpha$ toward the Heisenberg limit. As decoherence increases, the QFI becomes weakly dependent on layer count. For example, for $\gamma/\chi = 0.1$ and $0.25$ [see panels~\ref{fig:v-alpha}(b) and~\ref{fig:v-alpha}(c)], $n=3$ E--R layers already achieve nearly the same QFI as $n=7$ layers. 

For $\gamma/\chi = 0.01$ and $\gamma/\chi = 0.1$, the $1$-layer Ising configuration exhibits a sharply decreasing and non-monotonic dependence on $\alpha$~\cite{perlin2020spin, qfi_opt_manuscript1}. Increasing the layer count mitigates this rapid decline, although improvements at larger $\alpha$ are more gradual. Overall, increasing the layer count in the Ising model enhances the achievable QFI and drives a transition in the optimized states from squeezed-like to cat-like regimes. 

On the other hand, FTAT protocols [panels (d)--(f)] show limited benefits from layering, with improvements confined to a narrow range of $\alpha$ and rapidly suppressed by decoherence. For $\gamma/\chi = 0.01$ [panel (d)], the $1$-layer FTAT QFI decreases steadily with increasing $\alpha$ before recovering near $\alpha \approx 2.5$. Layering yields its largest gains for $\alpha \lesssim 3.6$ while providing only marginal improvements at larger $\alpha$. Notably, layering shifts the prominent QFI dip from the $1$-layer case at $\alpha \approx 2.5$ to $\alpha \approx 1.7$ for $n \geq 3$. 

At stronger decoherence, $\gamma/\chi = 0.1$ and $0.25$ [panels (e) and (f)], increasing the layer count produces only weak QFI enhancements. Nevertheless, even the $1$-layer FTAT protocol remains above the SQL across $\alpha$ and the decoherence considered. For FTAT overall, we observe rapid saturation with layer number $n$, with the $3$-layer implementation achieving QFIs comparable to the $7$-layer results.

While multilayer FTAT circuits modestly outperform their $1$-layer counterparts, the sensitivity to layering is appreciable primarily at $\gamma/\chi = 0.01$. This contrasts with the Ising model, for which layering yields systematic and robust gains across $\alpha$. 

\section{Analysis of system size}\label{sec:system-size}

\begin{figure*}
    \centering
    \includegraphics[width=0.48\linewidth]{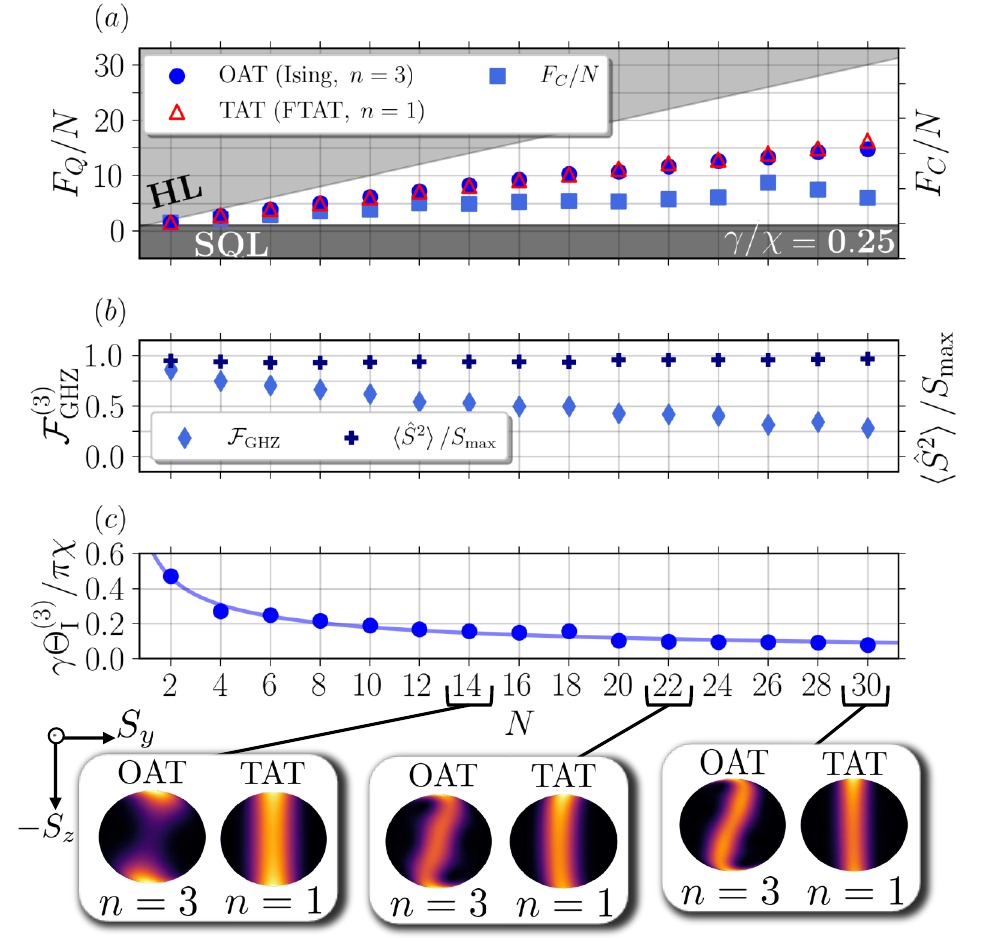}
    \includegraphics[width=0.48\linewidth]{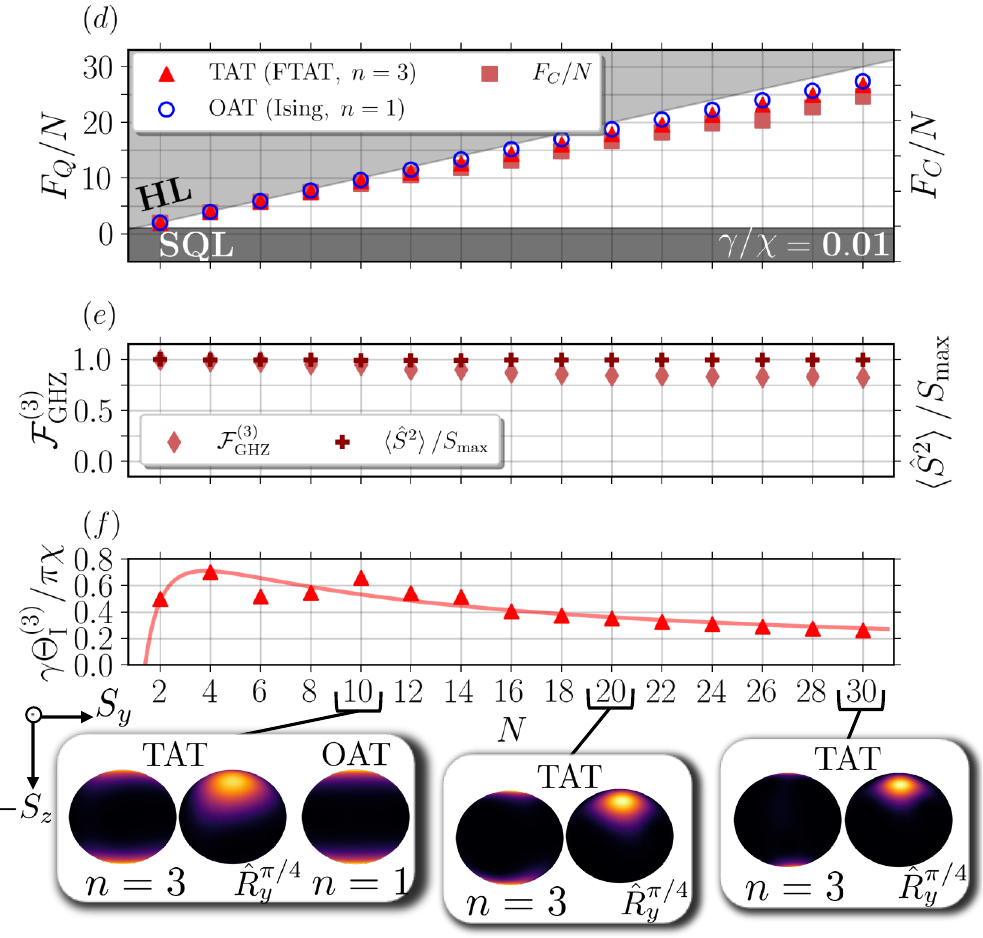}
    \caption{Diagnostic quantities as a function of system size $N$ for $3$-layer OAT at $\gamma/\chi = 0.25$ [(a)--(c)] and 3-layer TAT at $\gamma/\chi = 0.01$ [(d)--(f)]. In subplots (a)[(d)], we compute the QFI $F_Q/N$ (filled circles)[(filled triangles)] and CFI $F_C/N$ (squares)[(squares)] for generated states. For comparison, we overlay $1$-layer QFIs for TAT (open triangles) and OAT (open circles) in subplots (a) and (d), respectively. [(b) and (e)] State collectivities $\braket{\hat{S}^2}/S_{\mathrm{max}}$ (crosses) and GHZ state fidelities $\mathcal{F}^{(n)}_{\mathrm{GHZ}}$ (diamonds). 
    [(c) and (f)] Relative competition between the given noise rate and the total entangling gate application across $3$-layers, $\gamma\Theta_{\mathrm{I}}^{(3)}/\pi\chi$ with respect to system size, $N$. (Below) Husimi-like distribution comparisons for generated states with $1$- and $3$-layer VQC architectures. Brighter regions correspond to higher probability densities. States with a $\hat{R}_y^{\pi/4}$ designation depict the state rotated by an angle of $\pi/4$ about $y$.} 
    \label{fig:v_N_analysis}
\end{figure*}

To study state generation at system sizes beyond $N = 8$, we exploit the permutational symmetry of the collective OAT and TAT dynamics (respectively, Ising and FTAT with $\alpha = 0$). 
In both cases, we find that the CES ($\Theta_{\mathrm{I}}^{(n)}$) decreases with increasing $N$, consistent with the growing fragility of many-body coherences. The two models differ, however, in the character of the generated states. For OAT, cat-like structure degrades with increasing $N$, even as the QFI remains useful. By contrast, TAT sustains cat-like state generation.

\subsection{OAT}
In Fig.~\ref{fig:v_N_analysis}(a)--(c), we present diagnostic quantities for an OAT configuration as a function of system size $N$. For this case, we set $\gamma/\chi = 0.25$, which corresponds to the decoherence strength that yielded the greatest metrological enhancement for OAT in Sec.~\ref{sec:OAT}. Figure~\ref{fig:v_N_analysis}(a) depicts the QFI per particle, $F_Q/N$, for $3$-layer OAT (filled circles). The shaded regions mark the SQL (lower) and HL (upper) for reference. The $3$-layer OAT QFI consistently remains below the $1$-layer TAT (open triangles) QFI across $N$. 

We also compute and depict the state's CFI per particle [see Eq.~(\ref{eqn:cfi})] in Fig.~\ref{fig:v_N_analysis}(a) ($F_C/N$; squares). The CFI diverges progressively from the QFI as $N$ increases. The growing CFI-QFI disparity is consistent with local site dephasing and suggests that optimal readout increasingly relies on measurements of local observables as $N$ increases.

In Fig.~\ref{fig:v_N_analysis}(b) we plot the state collectivity ($\braket{\hat{S}^2}/S_{\mathrm{max}}$; $\bm{+}$) alongside the GHZ fidelity ($\mathcal{F}^{(3)}_{\mathrm{GHZ}}$; diamonds) for states generated with a $3$-layer OAT VQC. Across system sizes, the states remain highly collective but display a gradual decay in GHZ fidelity. Waning GHZ fidelity is consistent with loss of GHZ-like phase coherence when subjected to local noise, which decays roughly as $e^{-\gamma N t}$~\cite{Huelga_1997_decoherence, Foss-FeigM2013Dqco-ising-cat-states}. This coherence loss also helps explain the growing disparity between the QFI and the CFI. The distinction arises because the QFI quantifies the best achievable sensitivity over all possible measurements, whereas the CFI of Eq.~(\ref{eqn:cfi}) is constrained to collective readout. 

Figure~\ref{fig:v_N_analysis}(c) quantifies the interplay between decoherence and cumulative entangling duration by plotting their product $\gamma \Theta_{\mathrm{I}}^{(3)}/\pi \chi$ with respect to $N$. We observe that $\gamma \Theta_{\mathrm{I}}^{(3)}/\pi \chi$ decreases with increasing system size, consistent with the expectation that state coherences become more fragile at larger $N$. For this reason, we fit the numerical data to the empirical relation, $\gamma \Theta_{\mathrm{I}}^{(3)}/\pi\chi = aN^{-b}$, obtaining $a = 0.70 \pm 0.03$ and $b = 0.59 \pm 0.03$. The trend indicates that as $N$ increases, the optimized VQC reduces the total interaction time so that entangling dynamics occur before local decoherence substantially suppresses useful coherences. Nevertheless, while the CFI and $\mathcal{F}_{\mathrm{GHZ}}^{(3)}$ decrease for increasing $N$, the QFI is sustained above the SQL and consistently performs just below that achievable with $1$-layer TAT.

\subsection{TAT}
We now turn to the system-size analysis for $3$-layer TAT, shown in panels~\ref{fig:v_N_analysis}(d)--\ref{fig:v_N_analysis}(f), as a counterpart to the layered OAT results in panels~\ref{fig:v_N_analysis}(a)--\ref{fig:v_N_analysis}(c). For this case, we fix the decoherence to $\gamma/\chi = 0.01$, which corresponds to the decoherence strength yielding the largest QFI gain for TAT in Sec.~\ref{sec:TAT}. In Fig.~\ref{fig:v_N_analysis}(d) we show the QFIs of $1$-layer OAT (open circles) and $3$-layer TAT (filled triangles). We find that $3$-layer TAT sustains QFIs comparable to those of $1$-layer OAT across the system sizes surveyed, with the OAT configuration appearing to set an upper bound. The computed CFIs for $3$-layer TAT ($F_C/N$, squares) closely follow the corresponding QFI, indicating that collective readout protocols may exploit most of the state's metrological potential.

In Fig.~\ref{fig:v_N_analysis}(e) we similarly present the $3$-layer TAT state collectivity ($\bm{+}$) and GHZ fidelity (diamonds). Across the system sizes considered, we observe the generation of collective states marked with near-maximal collectivities, $\braket{\hat{S}^2}/S_{\mathrm{max}}\approx 1.0$. At the same time, the GHZ fidelity steadily decays with increasing $N$. The fidelity decay is consistent with the loss of state coherences, which scale roughly as $e^{-\gamma N t}$ and become increasingly fragile with system size. Nevertheless, the $3$-layer TAT VQC configuration sustains the generation of cat-like states for large $N$, as indicated by strong collectivities ($\braket{\hat{S}^2}/S_{\mathrm{max}}\approx 1.0$), appreciable GHZ fidelities ($\mathcal{F}^{(3)}_{\mathrm{GHZ}}\gtrsim0.75$) and Heisenberg-limited QFIs.

In Fig.~\ref{fig:v_N_analysis}(f) we show the product of the fixed decoherence and the CES [$\gamma \Theta^{(3)}_{\mathrm{I}}/\pi\chi$ with $\gamma/\chi = 0.01$] as a function of system size. We observe non-monotonic behavior for $N < 10$ and monotonically decreasing values for $N \geq 10$. Similar to the OAT case, we empirically fit $\gamma\Theta^{(3)}_{\mathrm{I}}/\pi\chi$. To capture the more nuanced behavior of the TAT case, we fit our results to $\log{(aN)}/bN$ and obtain the fit parameters $a = 0.71 \pm 0.05$ and $b = 0.37 \pm 0.02$. This choice of fit is motivated by two observations. First, the rate of TAT dynamics accelerates with increasing $N$, offering an increasing advantage against the set decoherence. Second, our results show that the time required to generate states follows asymptotic behavior similar to that observed in Refs.~\cite{spin-squeezing-ueda, PhysRevA.90.062132, tat-witkowska}.

\section{Conclusion} \label{sec:conclusion}
In this work, we developed and analyzed a layered VQC architecture inspired by twist-and-turn dynamics. Using experimentally motivated Hamiltonians in the presence of noise, we identified state-preparation strategies for generating nonclassical states with enhanced metrological utility.

For Ising-based Hamiltonians, layering systematically yields larger achievable QFIs across interaction ranges. For collective OAT and finite-range Ising, layering improves robustness to noise, in some cases through a reduction of cumulative entangling duration. On the other hand, for FTAT-based Hamiltonians, improvements are comparatively more limited as the interaction range decreases. Nevertheless, collective TAT remains a notable exception in which layering enables the preparation of GHZ states that are otherwise inaccessible under the native Hamiltonian alone. \Rev{These results demonstrate that layered architectures offer a robust and versatile framework for generating metrologically useful states in noisy settings and across diverse interaction regimes.}

\Rev{Having established optimal state preparation under noise with a layered E--R VQC, a natural next step is to analyze the corresponding readout performance. It is typically the case that extracting useful readouts from complex many-body states is a challenging task, especially in the presence of noise and in circumstances with a limited measurement pool. Several techniques address the latter challenge, particularly interaction-based readout protocols based on time-reversal~\cite{RevModPhys.90.035005, TnT-IBR-Haine, Haine_2018_IBR, RLS_2024_nonclassicalmotion} which can enable readouts with simpler or projective measurements~\cite{PhysRevA.94.010102, Liu2025Variational,  yicheng_dicke_2025}.

In future work, we may evaluate readout performance in noisy settings by embedding the states prepared in this manuscript in a time-reversal sensing protocol after encoding a collective $z$-rotation. The time-reversal stage specifically applies the state-preparation operations in reverse order with the Hamiltonian signs inverted, in the continued presence of decoherence. While we preliminarily expect noise during the time-reversal stage to further degrade the prepared state, and thus lessen the readout performance relative to the projected metrological potential, these results may be interesting for sensing applications with noisy intermediate-scale quantum (NISQ) devices~\cite{Jiao_2023}.}

\section*{Acknowledgments}
This work is supported by the U.S.~Department of Energy, Office of Science, Office of Advanced Scientific Computing Research, under the Exploratory Research for Extreme-Scale Science and accelerated Research for Quantum Computing programs. This material is also based upon work supported by the National Science Foundation under Grant No.~PHY-2110052. The authors gratefully acknowledge D.~Blume and K.~Mullen for their valuable feedback and insightful comments on early drafts of this manuscript. The majority of computing for this project was performed at the OU Supercomputing Center for Education \& Research (OSCER) at the University of Oklahoma (OU).

\bibliographystyle{apsrev4-2}
\bibliography{library}

\appendix
\section{State preparation strategies}\label{sec:state-prep-strategies}
In Sec.~\ref{sec:results}, we provided results generated with the Ising and FTAT Hamiltonians given by Eqs.~(\ref{eqn:ising-ham}) and (\ref{eqn:tat-ham}), respectively. There, we characterized the states found under infinite-range ($\alpha = 0$) and finite-range ($\alpha = 3$) interactions using diagnostic observables and aggregate quantities which were defined in Sec.~\ref{sec:diagnostic-tools}. In this section, we now present the corresponding state-preparation strategies explicitly by showing the QFI together with the optimized parameter set $\mathbf{x}_{\mathrm{opt}}^{(n)}$ as functions of $\gamma/\chi$.

\subsection{Ising-based VQCs}
\begin{figure*}
    \centering
    \includegraphics[width=0.48\linewidth]{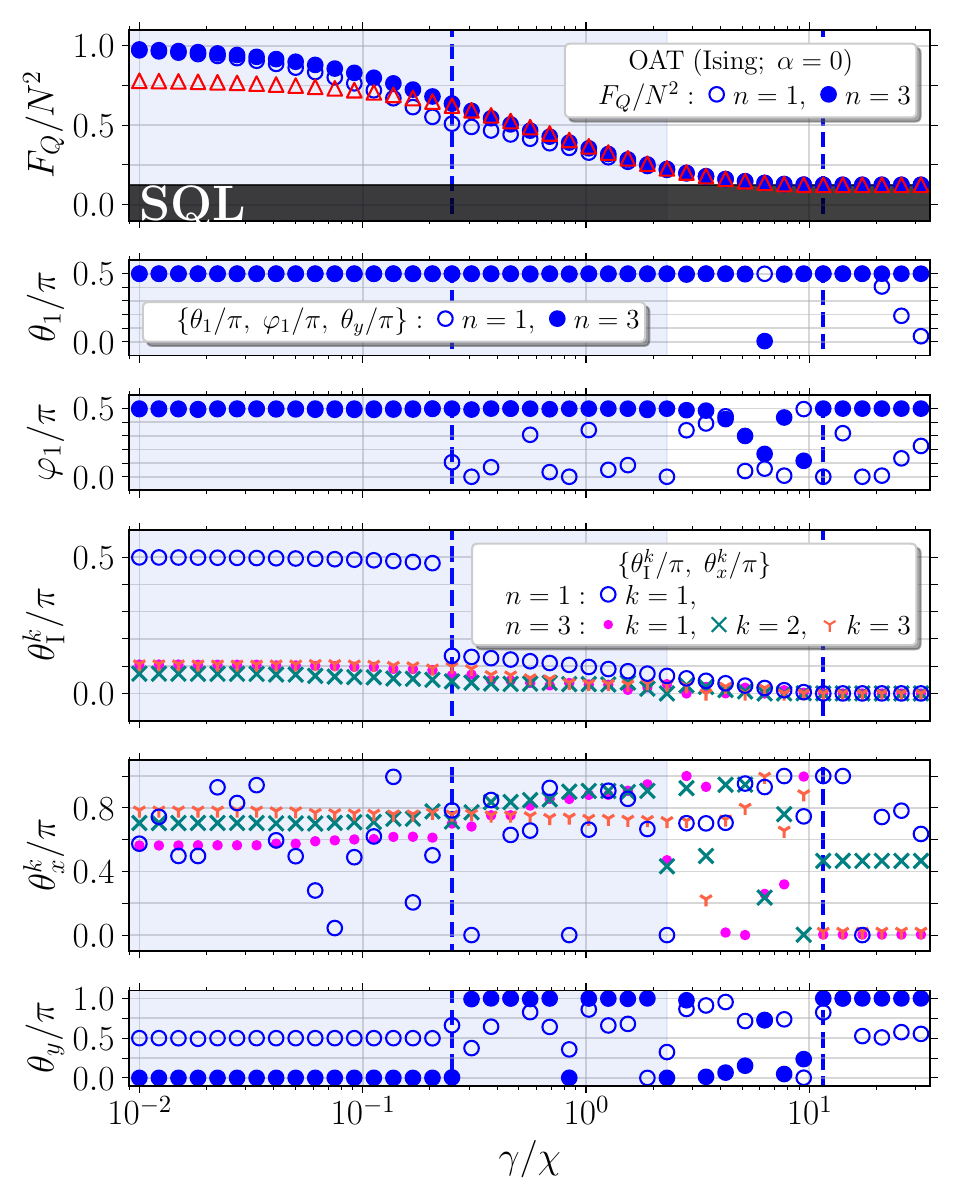}
    \includegraphics[width=0.48\linewidth]{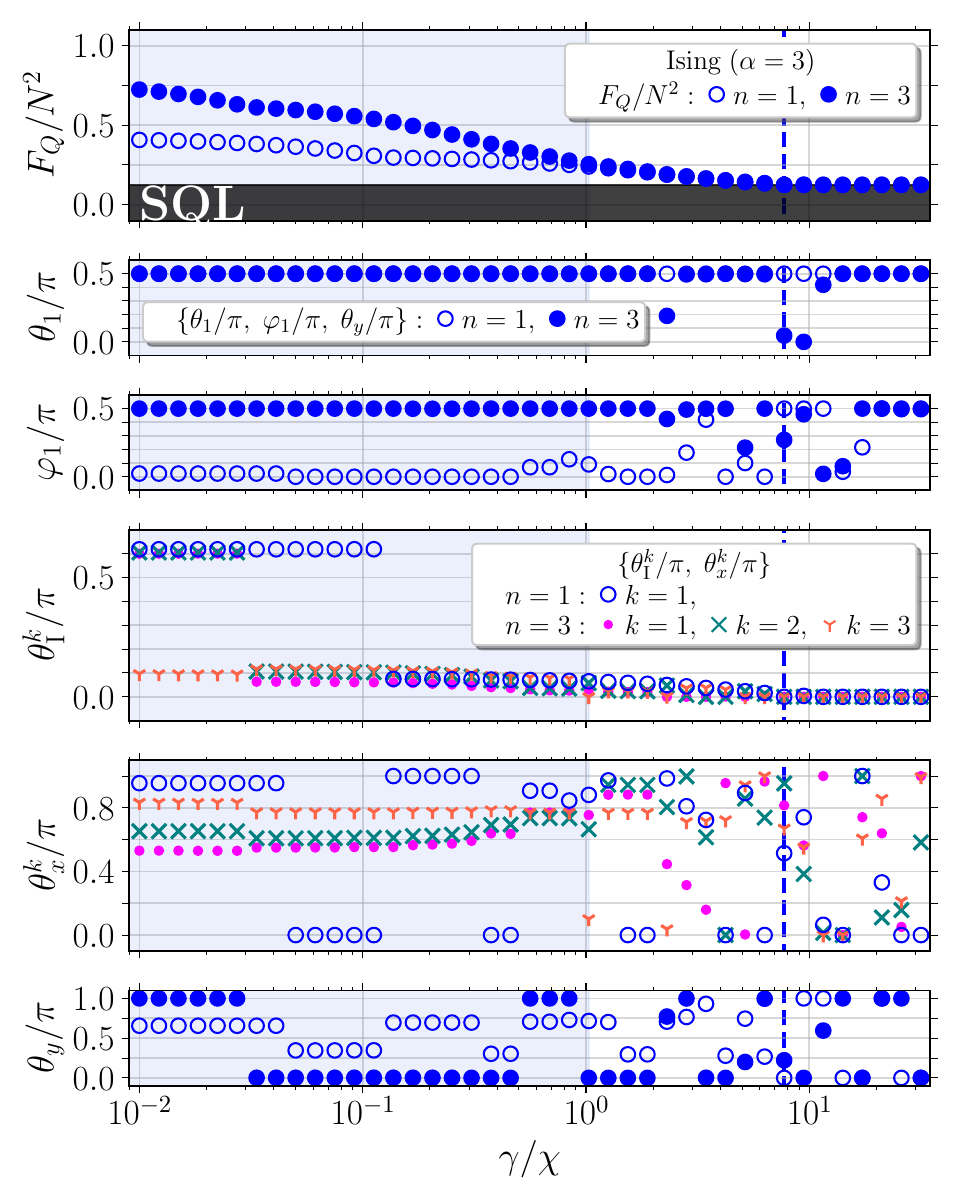}
    \caption{Optimal VQC parameter set $\mathbf{x}_{\mathrm{opt}}^{(n)}$ for OAT (Ising with $\alpha = 0$; left column) and Ising with $\alpha = 3$ (right column) for decoherence strength $\gamma/\chi$.}
    \label{fig:ising-param-set}
\end{figure*}

Figure~\ref{fig:ising-param-set} shows the preparation strategies for VQC implementations based on the Ising Hamiltonian [see Eq.~(\ref{eqn:ising-ham})]. In these results, the left (right) column corresponds to setting $\alpha = 0$ $(\alpha = 3)$. We find that OAT (Ising with $\alpha = 0$) and Ising with $\alpha = 3$ exhibit several similarities. For instance, in the $1$- and $3$-layer VQC of each case, the optimal first step is to rotate the initial state onto the Bloch equator with $\theta_1\approx \pi/2$ about the axis defined by $\varphi_1$. 

Once the state is rotated, the preparation strategy typically splits the cumulative entangling duration nearly evenly across the $n = 3$ layers with $\theta_{\mathrm{I}}^{k} \approx \Theta^{(3)}_{\mathrm{I}}/3$ for $k \in \{1, 2, 3\}$. This near-even splitting is observed for the majority of $\gamma/\chi$ in both cases, except for $\gamma/\chi < 0.02$ for the $\alpha = 3$ case, where the individual entangling strengths are approximately $\theta_{\mathrm{I}}^{1} = \theta_{\mathrm{I}}^{2}\approx 0.6 \pi$ and $\theta_{\mathrm{I}}^{3} \approx 0.1\pi$.

More nuanced splitting occurs between the three interleaved $x$-rotations ($\theta_x^k$). Over most of the MIR, subsequent $x$-rotations take on larger angles for increasing $k$ (i.e., $\theta_x^1 < \theta_x^2<\theta_x^3$). Moreover, each rotation tends to increase in line with a decreasing entangling strength $\theta_{\mathrm{I}}^{(k)}$. Beyond the MIR where QFI enhancement vanishes, the values of $\theta_x^{k}$ (and the CRA, $\Theta_x^{(3)}$) become discontinuous, serving only to reorient the state together with $\theta_y$.

Although the preparation strategies are similar for the $\alpha = 0$ and $\alpha = 3$ cases, the resulting QFIs and the nature of the generated states remain distinct. Nonetheless, for Ising interactions with $\alpha = 3$ (and in line with the observations of Sec.~\ref{sec:interaction-range}), increasing the layer number $n$ systematically improves performance and drives the achieved QFI closer to that of the collective $\alpha = 0$ case.

\subsection{FTAT-based VQCs}
\begin{figure*}
    \centering
    \includegraphics[width=0.48\linewidth]{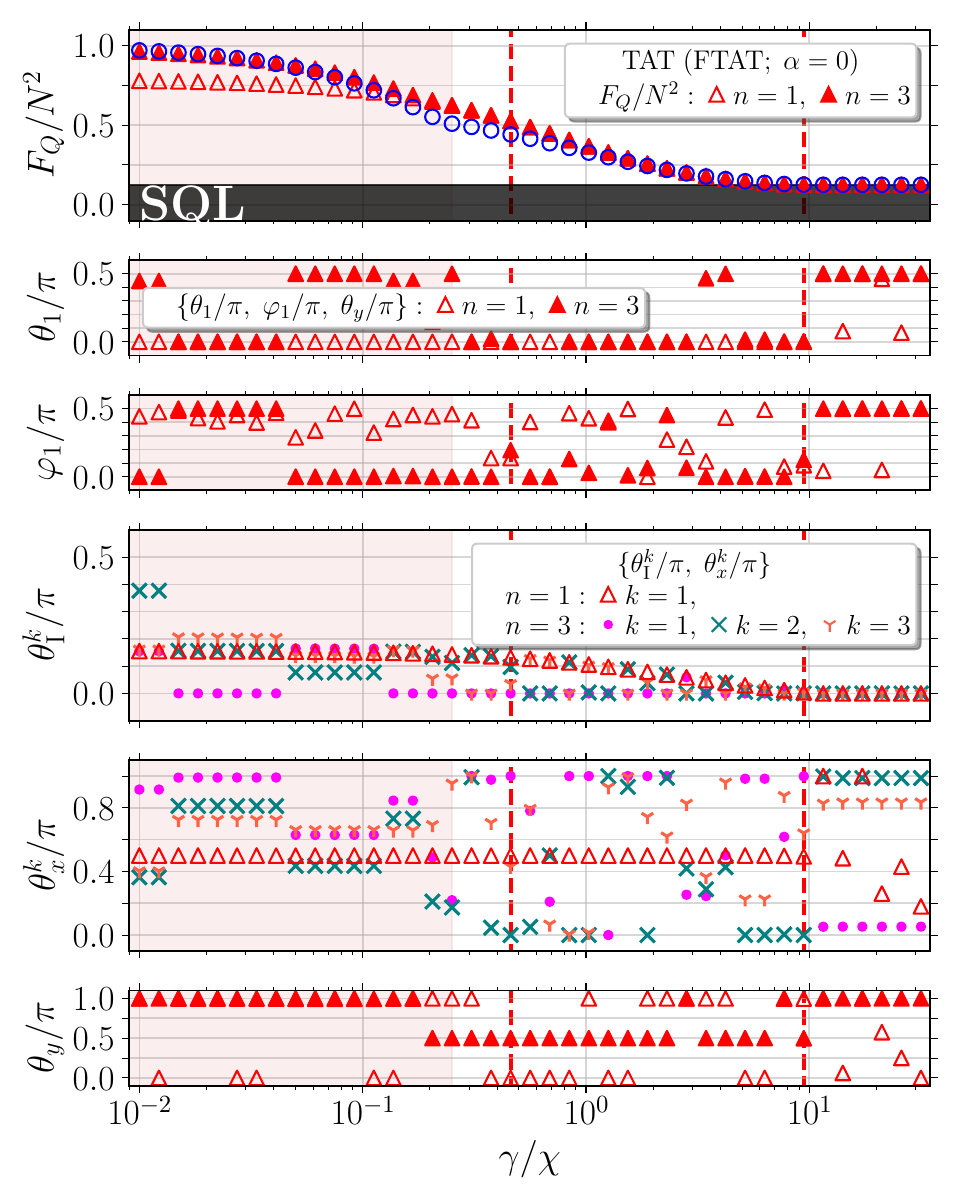}
    \includegraphics[width=0.48\linewidth]{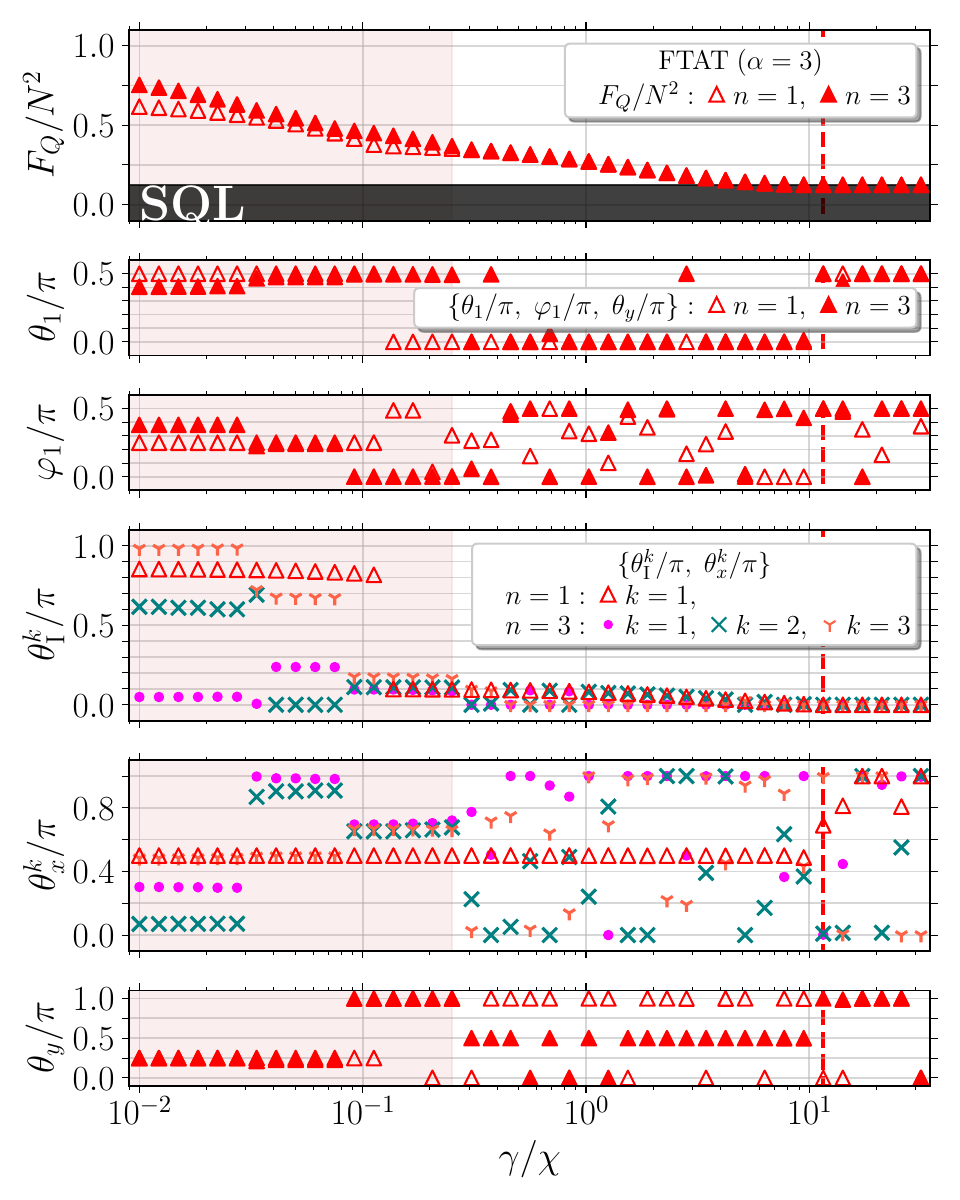}
    \caption{Optimal VQC parameter set $\mathbf{x}_{\mathrm{opt}}^{(n)}$ for TAT (FTAT with $\alpha = 0$; left column) and FTAT with $\alpha = 3$ (right column) for decoherence strength $\gamma/\chi$.}
    \label{fig:local_TAT-param-set}
\end{figure*}

We now show the optimal preparation strategies for FTAT VQCs [see Eq.~(\ref{eqn:tat-ham})]. In the left and right columns of Figure~\ref{fig:local_TAT-param-set}, we provide results for TAT (FTAT with $\alpha = 0$) and FTAT with $\alpha = 3$, respectively. In contrast to the results for Ising implementations presented in the last section, the optimal preparation strategies ($\mathbf{x}_{\mathrm{opt}}^{(n)}$) for FTAT with $\alpha = 0$ and $\alpha = 3$ are dissimilar.

Within the MIR, the initial rotations for TAT either rotate the state near the Bloch equator or do not rotate at all. On the other hand, FTAT with $\alpha = 3$ mainly rotates the state to the equator. Following the rotation, entangling strengths split unevenly across the $n = 3$ layers for both cases. We note that although the case of $\alpha = 3$ presents distinctly larger values of $\theta_{\mathrm{I}}^{(k)}$ relative to the $\alpha = 0$ case, the QFI increases only slightly. By contrast, significant QFI improvement occurs for $\alpha = 0$ despite having relatively weaker individual entangling strengths.

Similar to the individual entangling strengths, the interleaved $x$-rotations split unevenly across the MIR for both cases. More importantly, the optimized parameters exhibit plateaus across the MIR, particularly in $\theta_1$, $\varphi_1$, $\theta_{\mathrm{I}}^{(k)}$, and $\theta_x^{(k)}$. These plateaus are consistent with the persistence of similar optimized state-preparation strategies across broad ranges of $\gamma/\chi$. We further reconcile close similarity in the generated states with a smoothly varying QFI (and supplemental diagnostic tools).

\section{Driven Hamiltonians}\label{sec:driven_hams}
To gain insight into the effect that our VQC's layered E--R operations have on state generation, we consider an ideal case with OAT interactions in the unitary limit (i.e., with the Ising Hamiltonian of Eq.~(\ref{eqn:ising-ham}), setting $\alpha = 0$ and $\gamma/\chi = 0$). We assume that the VQC has infinite E--R layers ($n \to \infty$) and that the cumulative quantities $\Theta_{\mathrm{I}}^{(n)} = \chi t$ and $\Theta_x^{(n)} = \Omega t$ with coupling strength $\chi$ and an external drive strength $\Omega$ are equally divided by duration $t/n$ across layers. In this limit, the VQC's E--R sequence emulates a ``trotterized'' scheme~\cite{trotter, suzuki}
\begin{equation}
    e^{-i t(\chi \hat{S}_z^2 + \Omega \hat{S}_x)} = \lim_{n\to\infty}{(e^{-i \Theta_{\mathrm{I}}^{(n)} \hat{S}^2_z/n} \; e^{-i \Theta_x^{(n)}\hat{S}_x/n})^n},
\end{equation}
that describes evolution with the driven OAT Hamiltonian, 
\begin{equation}\label{eqn:driven-oat}
    \hat{H}_{\mathrm{I}} = \chi \hat{S}_z^2 + \Omega \hat{S}_x.
\end{equation}
A straightforward expansion of Eq.~(\ref{eqn:driven-oat}) using the collective raising/lowering spin operators $\hat{S}^{\pm} = \hat{S}_y \pm i \hat{S}_z$ gives
\begin{equation}\label{eqn:expanded-driven-OAT-ham}
    \hat{H}_{\mathrm{I}} = -\frac{\chi}{4}(\hat{S}^{+2} + \hat{S}^{-2}) - \frac{\chi}{2}\hat{S}_x^2 + \Omega\hat{S}_x,
\end{equation}
where we have discarded a collective term $\hat{S}^2$. Expression~(\ref{eqn:expanded-driven-OAT-ham}) is then expanded using linear stability analysis in which the state is assumed polarized along the $x$-axis so that $\hat{S}_x$ can be approximately written as a mean plus a small fluctuation, $\hat{S}_x \approx \braket{\hat{S}_x} + \hat{\delta}_x$. By setting $\Omega = \chi \braket{\hat{S}_x} / 2$, the expanded expression can be further reduced to the first term of Eq.~(\ref{eqn:expanded-driven-OAT-ham}). 

This final expression closely resembles the TAT Hamiltonian of Ref.~\cite{spin-squeezing-ueda} (e.g., $\hat{H} = \chi(\hat{S}^{+2}-\hat{S}^{-2})/2i$) and shows that for a properly chosen drive strength $\Omega$, a native OAT Hamiltonian can realize TAT-like behavior. From a complementary perspective, a driven TAT Hamiltonian may enable OAT-like behavior so long as the native TAT Hamiltonian is applied along the proper basis alongside the interleaved rotations. In our VQC, the initial rotation $\hat{R}_{\theta_1}^{\varphi_1}$ may select the appropriate basis.

From a pragmatic standpoint, infinite E--R layers are not required. For finite E--R layer treatments, the simulation of one Hamiltonian via another is approximate. For example, a single OAT interaction period interleaved with rotations has been shown to produce TAT-like dynamics to first order for sufficiently short durations~\cite{LiuYC2011Ssto}. 

\subsection{Toward cat-like state generation}
We had initially conjectured that $3$-layer OAT (TAT) VQC implementations would generate $1$-layer TAT-like (OAT-like) dynamics. This idea was supported by comparable QFIs obtained through $1$-layer and $3$-layer VQCs. In the case of $3$-layer OAT, the QFI approaches the QFI for $1$-layer TAT for $\gamma \gtrsim \gamma_1$ (see open triangles in Fig.~\ref{fig:ising-param-set}). Conversely, the QFI for $3$-layer TAT approaches the $1$-layer OAT QFI for $\gamma/\chi\lesssim10^{-1}$ (see open circles in Fig.~\ref{fig:local_TAT-param-set}). Similar characteristics are also seen at larger system sizes. For example, in Fig.~\ref{fig:v_N_analysis}(a) and~\ref{fig:v_N_analysis}(d), we find that layered OAT (TAT) implementations consistently achieve QFIs approaching their $1$-layer TAT (OAT) counterparts. Thus, we speculated whether layering under OAT (TAT) enabled the emulation of single-layer TAT (OAT) behavior.

While the QFIs of each case were closely related, optimal generation strategies favored cat-like states (decoherence permitting) as layering typically improved GHZ fidelities. This outcome is natural in the problem at hand because optimization tends toward the higher-QFI, cat-like class of states. Nevertheless, the generation of GHZ states with layered TAT does not preclude the emulation of OAT-like behavior.

\section{Husimi-like probability distributions}
\label{sec:husimi}

Here we define the probability distributions that we use to visualize spin states in the main text.
We note that this section is reproduced from Ref.~\cite{qfi_opt_manuscript1} for reference.
The basic idea is to define a Husimi probability distribution within each manifold of fixed spin length $S$ and take an appropriately weighted average of these distributions.
Note that, for ease of language, in this appendix we treat the symbol $S$ as a variable to denote spin length, whereas the main text defines $S=N/2$ for $N$ spins.

\subsection{A simplified case}

Consider first the simplified case in which $N$ spins occupy the permutationally symmetric manifold with spin length $S=N/2$.
In this case, for a density matrix $\hat\rho$ we can define the Husimi-like probability distribution
\begin{align}
  \rho(\bm v) = w_S \braket{\bm v|\hat\rho|\bm v}
\end{align}
\begin{align}
  \ket{\bm v} = \hat R(\bm v) \ket{\uparrow}^{\otimes N},
  &&
  \hat R(\bm v) = e^{-i\phi\hat S_z} e^{-i\theta\hat S_y}.
\end{align}
The normalization factor $w_S$ is determined by requiring that $\rho(\bm v)$ be a normalized probability distribution when integrated over the 2-sphere $\mathbb{S}_2$, with
\begin{align}
  \int_{\mathbb{S}_2} \text{d}\bm v\, \rho(\bm v)
  = \int_0^\pi \text{d}\theta\, \sin\theta
  \int_0^{2\pi} \text{d}\phi \, \rho(\theta,\phi)
  = 1,
\end{align}
which implies that\footnote{The normalization factor $w_S$ is most easily determined by considering a uniform mixture of all $N+1$ permutationally symmetric states, for which $\braket{\bm v|\hat\rho|\bm v}=1/(N+1)=1/(2S+1)$.}
\begin{align}
  w_S = \frac{2S+1}{4\pi}.
\end{align}
When the state $\hat\rho$ is permutationally symmetric, the probability distribution $\rho$ is a faithful representation of $\hat\rho$ in the sense that $\hat\rho$ is uniquely determined by $\rho$.
Moreover, in this case the values $\rho(\bm v)$ at $(2S+1)^2$ points $\bm v$ suffice to reconstruct $\hat\rho$~\cite{perlin2021spin}.

\subsection{The general case}

More generally, the state $\hat\rho$ may have components with non-maximal spin length $S$.
In this case, the state $\hat\rho$ can no longer be faithfully represented by a probability distribution on a sphere.
Nonetheless, we meaningfully visualize the spin polarization of $\hat\rho$ by averaging over probability distributions that represent components of $\hat\rho$ within fixed-$S$ manifolds.
To this end, we classify states by their spin length $S$, spin projection $m$ onto a quantization axis, and an auxiliary index $\xi$ that encodes how a state transforms under spin permutations.
That is, we identify states $\ket{S,m,\xi}$ that are simultaneous eigenstates of $\hat{S}^2 = \hat S_x^2 + \hat S_y^2 + \hat S_z^2$ and $\hat S_z$, with $\hat{S}^2\ket{S,m,\xi}=S(S+1)\ket{S,m,\xi}$ and $\hat{S}_z\ket{S,m,\xi}=m\ket{S,m,\xi}$.
We then define the rotated state
\begin{align}
  \ket{\bm v,S,\xi} = \hat R(\bm v) \ket{S,S,\xi},
\end{align}
which is the analogue of $\ket{\bm v}$ in a sector of Hilbert space with spin length $S$.
The net spin-polarization probability distribution $\rho(\bm v)$ is then
\begin{align}
  \rho(\bm v) = \sum_{S,\xi} \, w_S \braket{\bm v,S,\xi|\hat\rho|\bm v,S,\xi}.
  \label{eq:distribution_def}
\end{align}

\section*{Disclaimer}
This paper was prepared for informational purposes by the Global Technology Applied Research center of JPMorgan Chase \& Co. This paper is not a product of the Research Department of JPMorgan Chase \& Co.~or its affiliates. Neither JPMorgan Chase \& Co.~nor any of its affiliates makes any explicit or implied representation or warranty and none of them accept any liability in connection with this paper, including, without limitation, with respect to the completeness, accuracy, or reliability of the information contained herein and the potential legal, compliance, tax, or accounting effects thereof. This document is not intended as investment research or investment advice, or as a recommendation, offer, or solicitation for the purchase or sale of any security, financial instrument, financial product or service, or to be used in any way for evaluating the merits of participating in any transaction.

\end{document}